\documentclass[preprint,12pt]{elsarticle}

\usepackage{amsmath,amsfonts,amssymb}
\usepackage{nomencl}
\usepackage{makecell}
\usepackage{mathtools}
\usepackage{qcircuit}
\usepackage{bbding}
\usepackage{array}
\usepackage{braket}
\usepackage{hyperref}
\usepackage{enumitem}
\usepackage{epstopdf}
\usepackage{amsthm}
\usepackage{mathpazo}
\usepackage{array}
\usepackage{geometry}
\usepackage{slashed}
\usepackage{tikz}
\usepackage{CJK}
\usepackage[caption=false,font=normalsize,labelfont=sf,textfont=sf]{subfig}
\usepackage{textcomp}
 \usepackage{natbib}
\usepackage{stfloats}
\usepackage{enumerate}
\usepackage{makecell}
\usepackage{multirow}
\usepackage{enumitem}
\usepackage{url}
\usepackage{verbatim}
\usepackage{graphicx}
\usepackage{tabularx}
\usepackage{algorithm}
\usepackage{algorithmicx}
\usepackage{algpseudocode}
\usepackage{etoolbox}

 \usepackage{lineno}
\newtheorem{definition}{Definition}[section]
\newtheorem{example}{Example}

\newcommand{\tabincell}[2]{\begin{tabular}{@{}#1@{}}#2\end{tabular}}
\newcommand{\circledsmall}[1]{\hbox{\tikz\draw (0pt, 0pt)
    circle (.45em) node {\makebox[0.15em][c]{\scriptsize#1}};}}
\newcommand{\circledtiny}[1]{\hbox{\tikz\draw (0pt, 0pt)
    circle (.3em) node {\makebox[0.15em][c]{\tiny#1}};}}

 \usepackage{url} 

\journal{LATEX}

\begin{document}

\begin{frontmatter}



\title{BF-QC: Belief Functions on Quantum Circuits}


\author[inst1]{Qianli Zhou}

\affiliation[inst1]{organization={Institute of Fundamental and Frontier Science},
organization={UESTC},
            city={Chengdu},
            postcode={250014},
            country={China}}

\author[inst3]{Guojing Tian}
\affiliation[inst3]{organization={Institute of Computing Technology},
organization={CAS},
            city={Beijing},
            postcode={100086},
            country={China}}

\author[inst1,inst2,inst4]{Yong Deng}

\affiliation[inst2]{organization={School of Education},
organization={SNNU},
            city={Xi'an},
            postcode={710062},
            country={China}}

\affiliation[inst4]{organization={Department of Management, Technology, and Economics},
organization={ETH Zurich},
            city={Zurich},
            postcode={8093},
            country={Switzerland}}

\begin{abstract}
Dempster-Shafer Theory (DST) of belief function is a basic theory of artificial intelligence, which can represent the underlying knowledge more reasonably than Probability Theory (ProbT). Because of the computation complexity exploding exponentially with the increasing number of elements, the practical application scenarios of DST are limited. In this paper, we encode Basic Belief Assignments (BBA) into quantum superposition states and propose the implementation and operation methods of BBA on quantum circuits. We decrease the computation complexity of the matrix evolution on BBA (MEoB) on quantum circuits. Based on the MEoB, we realize the quantum belief functions' implementation, the similarity measurements of BBAs, evidence Combination Rules (CR), and probability transformation on quantum circuits.
\end{abstract}


\end{frontmatter}


\section{Introduction}
\label{sec:sample1}

For a random variable with $n$ basic events, the classical Probability Theory (ProbT) uses $n$-dimensional Probability Mass Function (PMF) as a basic unit to represent information. However, in practice, we may lack information to describe the information as an accurate PMF. Based on the multi-valued probabilitic mapping, Dempster \cite{dempster2008upper} and Shafer \cite{shafer1976mathematical} propose Dempster-Shafer Theory (DST). Evidence Combination Rules (CR) and belief functions in DST have stronger ability to model and handle uncertain information than PMF, and are widely applied to information fusion \cite{yang2013evidential,Huang2021Cross,bronevich2021measures}, decision-making \cite{yang2002nonlinear,fei2022optimzation}, and reliability analysis \cite{mi2018reliability,xiahou2022fusing}. Smets \cite{smets1994transferable} introduces open-world assumption in DST and proposes the Transferable Belief Model (TBM), which argues that the set-valued weights represent ignorance. Based on the 'imprecise' and 'randomness', DST is used to strengthen the robustness of machine learning algorithms such as evidential neural network \cite{denoeux2019logistic}, classifiers \cite{liu2019evidence,8878018} and decision trees \cite{ma2016online}.

Though DST is widely used in uncertain information processing, there are still three issues that restrict the development of DST into a more general artificial intelligence approach. The first issue is how to merge evidential information reasonably under the uncertain environments. This topic has been discussed at length based on the handling of conflicting evidence and the Dempster Rule of Combination (DRC). In the recent past, several new fusion scheme is proposed to provide new perspectives for this topic. Pichon \textit{et al.} \cite{pichon2019quality} introduce relevance and truthfulness to propose meta-knowledge to correct and fuse information. Dubois \textit{et al.} \cite{dubois2020prejudice} utilize diffidence function to introduce the prejudice in infromation fusion. Bronevich  and Lepskiy \cite{bronevich2021measures} propose a new measure of internal and external conflict, and resolve the conflict based on the clustering. The second issue is how to model the total uncertainty of belief functions \cite{Deng2020ScienceChina}. Scholars have tried many methods to solve this problem, such as Shannon entropy \cite{jousselme2006measuring}, generalized Hartley entropy \cite{jirouvsek2018new}, belief interval \cite{wang2018uncertainty,yang2016new}, and process of Pignistic Probability Transformation (PPT) \cite{zhou2022higher}. Requirements \cite{abellan2008requirements} and evaluations \cite{abellan2017critique,abellan2016drawbacks} of these methods are also proposed. But up to now, there is no method that can not only satisfy all requirements but measure total uncertainty reasonably. The last and most important issue is that the computational complexity of the algorithms of belief functions explode exponentially with the increasing number of elements. For the evidence CRs, Tessem \cite{tessem1993approximations} utilizes an approximate method to sparse BoE to reduce the computation complexity, Benalla \textit{et al.} \cite{benalla2021computational} utilize parallel computing, and Rico \textit{et al.} utilize native GPU implementation, both to achieve the efficient CR algorithm. Chaveroche \textit{et al.} \cite{chaveroche2019efficient} propose a fast M$\mathrm{\ddot{o}}$bius transformation and applied it in DST. For evidential reasoning, Jean and Edward \cite{Jean1985hierarchical} model belief functions on a hierarchical hypothesis space to simplify its calculation process. For uncertainty measure, Liu \textit{et al.} \cite{liu2007reducing} propose a method to accelerate the calculation of the Ambiguity measure. However, the above methods all give corresponding strategies for special algorithms, rather than an efficient unified model.

In addition to DST, more and more artificial intelligence algorithms also face the problem of high computational complexity caused by excessive data volume. Quantum computers, as one of the solutions to deal with the limit of Moore's Law, have widely been researched in recent years. Shor decomposition algorithm \cite{shor1999polynomial} can decompose large prime numbers on quantum computers in polynomial time, but the same operation requires exponential time complexity on classical computers. The global search algorithm proposed by Grover  \cite{grover1996fast} can also achieve square-level acceleration on quantum computers. These algorithms prove that quantum computers have significant advantages over classical computers in certain problems, but due to the special requirements \footnote{state preparation and unitary operation, etc.} of quantum computing, quantum algorithms do not have much development in the field of machine learning. Until 2008, Giovannetti \textit{et al.} propose Quantum Random Access Memory (QRAM) and its implementation method \cite{giovannetti2008architectures}, and in 2009, Harrow \textit{et al.} propose the HHL algorithm \cite{harrow2009quantum}, which achieves the exponential acceleration on quantum computers to solve Linear System Problems (LSP). The above two works greatly promote the development of Quantum Machine Learning (QML) \cite{biamonte2017quantum}. In recent years, more and more quantum machine learning algorithms have been proposed, such as quantum K-means cluster \cite{lloyd2013quantum}, quantum Bayesian network \cite{low2014quantum}, quantum principle component analysis \cite{lloyd2014quantum}, quantum neural networks \cite{schuld2014quest}. Compared with classical computers, quantum computers use $n$ qubits to represent $2^n$ data by encoding them into superposition states, which has mathematical consistency with DST \footnote{most $2^n$ focal sets under an $n$-element framework}.

There has been previous work on the use of quantum logic to describe DST. Resconi and Nikolov \cite{resconi2001tests} model interference experiments using mass functions of focal sets. Vourdas \cite{vourdas2014quantum,vourdas2014lower} uses the non-additivity of quantum probabilities to model the mass functions of DST in lattice subspaces, which focuses on the consistency of physical meaning of lattice subspaces and elements. Xiao \cite{xiao2021caftr,Xiao2022NQMF} proposes the complex-valued DST and utilized the interference effect in quantum-like decision-making to simulate the human personality, but it is only a mathematical extension without realization on quantum computers. Deng and Jiang \cite{deng2021quantum} represent BBA by quantum mixed states, whose model has consistency with pignistic probability transformation but lacks implementation methods and application fields. Pan and Deng \cite{pan2020quantum} propose the representation method of Dempster Rule of Combination (DRC) in quantum states, but did not give a reasonable implementation method and reduce the complexity of the algorithm. Deng \textit{et al.} \cite{deng2020novel} propose a quantum-like model for information fusion. To sum up, none of the previous works used quantum computing to accelerate the DST algorithm, nor do they provide a method to implement belief functions on quantum circuits.

In this paper, we encode Basic Belief Assignment (BBA) into the amplitude of quantum superposition state to define the BBA on quantum circuits (BBA-QC), and combine QRAM to propose its implementation method on the quantum circuit. By corresponding $1$ element to $1$ qubit, we use C-NOT gates to capture the belief to implement belief functions on quantum circuits. In addition, we extend common operations of DST on quantum circuits and discuss their advantages in complexity. The structure of paper is shown as follows: Section \ref{pre} introduces the preliminary knowledge required for the paper, including the basic concepts of DST and quantum computing. Section \ref{3} defines the BBA-QC and implements it on quantum circuits. For belief functions, we use $C-NOT$ gates to capture the belief functions and simulate them on the Qiskit platform. In Section \ref{4}, we implement Matrix Evolution on BBA (MEoB) on quantum circuits and utilized it to handle BBA-QC. Section \ref{con} summarized the whole paper and discuss the future research directions.

\section{Preliminary}
\label{pre}

\subsection{Dempster-Shafer theory of belief functions}

We give the basic concepts of DST and its common operations.

\subsubsection{Basic Belief Assignment}

Dempster \cite{dempster2008upper} and Shafer \cite{shafer1976mathematical} propose the Basic Probability Assignment (BPA) to represent the underlying knowledge of PMF, and in TBM, Smets \cite{smets1994transferable} uses the Basic Belief Assignment (BBA) to emphasize the belief rather than probability. In this paper, BPA and BBA are considered to be the same.
For a frame of discernment (FoD) $\Theta=\{\theta_{1},\dots,\theta_{n}\}$, all mutually exclusive elements in $\Theta$ form a closed space. Its power set is written as $2^{\Theta}=\{\{\emptyset\},\{\theta_{1}\},\dots,\{\theta_{n}\},\{\theta_{1}\theta_{2}\},\dots,\{\theta_{1}\dots\theta_{n}\}\}$. BBA, also called mass function $m$, is defined on $2^{\Theta}$, and $m$ satisfies $m(F)\in[0,1]$ and $\sum_{F\subseteq\Theta}m(F)=1$. If $m(F)\neq0$, $F$ is the focal set. For convenience of representation, some special BBAs are displayed.

\begin{itemize}
\item \textbf{subnormal BBA:} Iff $m(\emptyset)\neq0$, $m$ is subnormal, and $m(\emptyset)$ represents the belief in open-world.
\item \textbf{Bayesian BBA:} Iff all focal sets are singletons, $m$ is Bayesian BBA, which is same with PMF.
\item \textbf{vacuous BBA:} Iff $m(\Theta)=1$, $m$ is vacuous BBA and represents total ignorance to $\Theta$.
\item \textbf{consonant BBA:} Iff all focal sets satisfies $F\subset G$ or $G \subset F$, $m$ is consonant BBA, and can be transformed to possibility distribution.
\end{itemize}

\subsubsection{belief functions}

belief ($Bel$) function, plausibility ($Pl$) function, and commonality ($q$) function are also basic information units, which have the same information content as BBA, and they can be transformed to each other through invertible matrices \cite{smets2002application}.
For a $m$ under an $n$-element FoD $\Theta$, its belief functions are defined as

\begin{equation}\label{bele}
Bel(F)=\sum_{\emptyset \neq G\subseteq F } m(G)=1-Pl(\overline{F});
\end{equation}

\begin{equation}\label{ple}
Pl(F)=\sum_{G \cap F \neq\emptyset } m(G)=1-Bel(\overline{F});
\end{equation}

\begin{equation}\label{qe}
q(F)=\sum_{F \subseteq G } m(G) ~
\mathrm{and}~m(F)=\sum_{F \subseteq G } (-1)^{|G|-|F|} q(F);
\end{equation}

where $\overline{F}$ means the complement of $F$, and the Eq.(\ref{qe}) is M$\mathrm{\ddot{o}}$bius transformation. $Bel$ function and $Pl$ function compose the belief interval $[Bel(F),Pl(F)]$ to represent the uncertainty of elements in focal set $F$.

\subsubsection{evidence Combination Rules}

In TBM, evidence CRs are used to transfer belief between focal sets, which is seen as a process of information fusion. Disjunctive Combination Rule (DCR) $\circledsmall{$\cup$}$ and Conjunctive Combination Rule (CCR) $\circledsmall{$\cap$}$ are defined as

\begin{equation}\label{ccr1}
m_1\circledsmall{$\cup$} m_2=m_{1\circledtiny{$\cup$}2}(F)=\sum_{F=G\cup H}m_1(G)m_2(H),
\end{equation}\label{dcr1}
\begin{equation}
m_1\circledsmall{$\cap$} m_2=m_{1\circledtiny{$\cap$}2}(F)=\sum_{F=G\cap H}m_1(G)m_2(H);
\end{equation}

and DCR and CCR also can be calculated by belief functions

\begin{equation}\label{dcr}
b_1\circledsmall{$\cup$} b_2=b_{1\circledtiny{$\cup$}2}(F)=b_1(F)\cdot b_2(F),
\end{equation}

\begin{equation}\label{ccr}
q_1\circledsmall{$\cap$} q_2=q_{1\circledtiny{$\cap$}2}(F)=q_1(F)\cdot q_2(F);
\end{equation}
where $b(F)=Bel(F)+m(\emptyset)$. DRC $\oplus$ as the core algorithm of DST, can be derived from normalized CCR
\begin{equation}\label{DRC}
m_1\oplus m_2=m_{1\oplus2}(F)=
\begin{cases}
0 & \text{$F=\emptyset$}\\
\frac{m_{1\circledtiny{$\cap$}2}(F)}{(1-m_{1\circledtiny{$\cap$}2}(\emptyset))} & \text{$F\neq\emptyset$}
\end{cases}.
\end{equation}
In \cite{smets2002application}, evidence CRs can be realized by the matrices operations.

\subsubsection{probability transformation}
In the decision layer of TBM, when no more information can be fused, BBAs are transformed into probability distributions to make decisions. PPT \cite{smets1994transferable} is the most common probability transformation method. Given a BBA $m$ under $\Theta$, the probability after transformation $BetP$ is $BetP(\theta_i)=\sum_{\theta_i\in F}\frac{m(F)}{|F|(1-m(\varnothing))}$. In addition, Cobb and Shenoy \cite{cobb2006plausibility} propose the Plausibility Transformation Method (PTM), which is the only transformation method that satisfies combination consistency\footnote{For $2$ BoEs, the order in which PTM and DRC are performed does not affect its outcome}. $Pl\_P$ is denoted as $Pl\_P(\theta_i)=\frac{Pl(\theta_i)}{\sum_{\theta_i\in\Theta}Pl(\theta_i)}$.

\subsubsection{total uncertainty of BBAs}

From the perspective of discord and non-specificity, Jirou{\v{s}}ek and Shenoy's (JS) entropy \cite{jirouvsek2018new} and Fractal-based Belief (FB) entropy \cite{zhou2022fractal} can measure the total uncertainty of BBA intuitively. Given a BBA $m$ under $n$-element FoD $\Theta$, JS entropy is derived from the sum of discord and non-specificity

\begin{equation}
\begin{aligned}
H_{\mathrm{JS}}(m)=H_{Pl\_P}+GH=-\sum_{\theta_i\in\Theta}Pl\_P(\theta_i)\log Pl\_P(\theta_i)+\sum_{F\subseteq\Theta}m(F)\log |F|,
\end{aligned}
\end{equation}

where $H_{Pl\_P}$ is Shannon entropy of $Pl\_P$ and $GH$ is the generalized Hartley entropy. And FB entropy is defined as

\begin{equation}
\begin{aligned}
H_{\mathrm{F}}(m)=-\sum_{F\subseteq\Theta}\sum_{F\subseteq G}\frac{m(G)}{2^{|G|}-1}\log\sum_{F\subseteq G} \frac{m(G)}{2^{|G|}-1}=-\sum_{F\subseteq\Theta}m_{\mathrm{F}}(F)\log m_{\mathrm{F}}(F),
\end{aligned}
\end{equation}
where $m_{\mathrm{F}}(F)$ called Fractal-based Basic Belief Assignment (FBBA) is a specialization of $m$ and satisfies $m\sqsubseteq_{s} m_{\mathrm{FB}}$. FBBA assigns ignorance of focal sets to their power sets, which can be seen as a transition of $m$ and represent a larger uncertainty than a uniform distribution. Hence, FB entropy can not only represent the total uncertainty of BBA when discord and non-specificity are linearly added, but settle the counter-intuitive results of Deng entropy \cite{abellan2017analyzing,moral2020critique} as well.

\subsubsection{Distance of BBAs}\label{dbba}

Characterizing the relationship between the evidence is beneficial to obtain more accurate results in practical applications. Since focal sets are not completely mutually exclusive, traditional distance measurement methods are not suitable for DST. Jousselme \textit{et al.} \cite{jousselme2001new} utilize a coefficient matrix $\boldsymbol{D_\mathrm{E}}$ propose the distance between BBAs. Given BBAs $m_1$ and $m_2$, the form of vector is $\boldsymbol{m'}=\{m(\theta_1)\cdots m(\Theta)\}^{T}$, and the evidence distance is

\begin{equation}\label{ed1}
d_\mathrm{BBA}(\boldsymbol{m_1'},\boldsymbol{m_2'})=\sqrt{\frac{1}{2}(\boldsymbol{m_1'}-\boldsymbol{m_2'})^T{\boldsymbol{D_\mathrm{E}}}(\boldsymbol{m_1'}-\boldsymbol{m_2'})},
\end{equation}
where matrix is a Hermitian matrix $\boldsymbol{D_\mathrm{E}}(F,G)=\frac{|F\cap G|}{|F\cup G|}$, and $d_\mathrm{BBA}\in[0,1]$.

The above parts are the basic concepts and common algorithms of DST. The following paper will discuss their implementation and operation on quantum circuits.

\subsection{Quantum computation}

We only introduce the basic operation rules on quantum circuits. For the specific corresponding physical process, please refer to \cite{nielsen2002quantum}.

\subsubsection{qubit}

The classical bit can only represent a definite state of $0$ or $1$, qubit represents a superposition state between $0$ and $1$, and only collapse to $0$ or $1$ when measured. In quantum computing, the matrix calculation is used to simulate physical processes, and the corresponding notations are shown in Table \ref{notation}.
\begin{table}[htbp!]
\caption{The description of mathematical notations in quantum computing}
\label{notation}
\begin{tabular}{c|c}
\hline
Notation & Mathematical Description \\
\hline
$\ket{\psi}$ & A complex-valued unit column vector\\
$\braket{\psi|\phi}$ & Inner product of $\ket{\psi}$ and $\ket{\phi}$\\
$A^*$ & Complex conjugate of matrix $A$\\
$H$ & Hermitian matrix, satisfy $H=H^\dagger$\\
\hline
Notation & Mathematical Description \\
\hline
$\bra{\psi}$ & A dual vector of $\ket{\psi}$\\
$\ket{\psi}\bigotimes \ket{\phi}$ & Tensor product of $\ket{\psi}$ and $\ket{\phi}$\\
$A^\dagger$ & Hermitian conjugate of matrix $A$, $A^\dagger=(A^T)^*$\\
$U$ & Unitary matrix, satisfy $UU^\dagger=I$\\
\hline
\end{tabular}
\end{table}

Quantum computing is operating quantum states in the Hilbert space, which is composed of mutually orthogonal vectors $\ket{\psi}$. A group of $\ket{\psi}$ is called the standard orthonormal basis, and under different basis, the state has different amplitudes. For a qubit, $\ket{0}$ and $\ket{1}$ are used as basis to represent a quantum superposition state

\begin{equation}
\ket{0}=\begin{bmatrix} 1  \\ 0 \end{bmatrix} ; \ket{1}=\begin{bmatrix} 0 \\ 1 \end{bmatrix} ; \ket{\Psi}=c_{1}\ket{0}+c_{2}\ket{1}=\begin{bmatrix} c_{1} \\ c_{2} \end{bmatrix},
\end{equation}
where $c_{1}$ and $c_{2}$ are complex number and satisfy $||c_{1}||^{2}+||c_{2}||^{2}=1$. $||c_{i}||^{2}$ is the probability amplitude of the state $\ket{i}$, which can be observed approximately after numerous measurements. For multiple qubits, tensor product $\bigotimes$ connect different states to form a new superposition state. Suppose two states $\ket{\Psi_{1}}=c_{1}\ket{0}+c_{2}\ket{1}$ and $\ket{\Psi_{2}}=d_{1}\ket{0}+d_{2}\ket{1}$, their joint state $\ket{\Psi_{1}}\bigotimes\ket{\Psi_{2}}=c_{1}d_{1}\ket{00}+c_{1}d_{2}\ket{01}+c_{2}d_{1}\ket{10}+c_{2}d_{2}\ket{11}$($\Psi_1\Psi_2$ briefly).

For a pure state, it can be described as a point on the Bloch sphere \cite{nielsen2002quantum}. By rotating $\phi$ around the $y$-axis and rotating $\theta$ around the $z$-axis, its amplitude and phase can be determined respectively. So given parameter $(\phi,\theta)$, any qubit can be written as $\ket{\Psi}=\cos (\frac{\theta}{2})\ket{0}+e^{i\phi}\sin (\frac{\theta}{2})\ket{1}$. For the mixed state, it is a point in the Bloch sphere, which is usually represented by a density matrix. Suppose in a quantum system, there are quantum state $\ket{\Psi_{i}}$ with the probability $p_{i}$, and the density matrix of it can be represented as $\rho=\sum_{i}p_{i}\ket{\Psi_{i}}\bra{\Psi_{i}}$.

\subsubsection{quantum gates}

Different from classical computing, quantum computing utilizes unitary transform on the state to generate the goal state, and capture the classic data by measurements. Quantum gate as an operator is expressed as a unitary matrix, For a quantum state with $n$ qubits, the operator is a $2^{n}\times2^{n}$ unitary matrix. For example, if a quantum gate
$
\Qcircuit @C=1em @R=.7em {
& \gate{U} & \qw
}$
operate on $\ket{\Psi_{0}}$, it can be written as $\ket{\Psi_{1}}=U\ket{\Psi_{0}}$. Because only unitary matrices can be implemented on quantum circuits, when executing a classical algorithm on a quantum circuit, first find the required unitary matrices and then implement them using basic gates.
Table \ref{s2t1} shows several common basic gates in quantum algorithms and their matrix representations. Given a quantum bit $\ket{\Psi}=c_1\ket{\psi}+c_2\ket{\phi}$, the gates can realize following operations: (1)Pauli-X gate is to reverses the state, i.e. $X\ket{0}\rightarrow \ket{1}$. (2)Y-Rotation gate is to rotate the state $\theta$ around the y-axis on Bolch sphere, which controls the amplitude of the state. (3)Z-Rotation gate is to rotate the state $\lambda$ around the z-axis on Bolch sphere, which is equivalent to adding a global phase to the $\ket{1}$ and maintaining the $\ket{0}$. (4)Hadamard gate is one of the most useful gates. It can transform $\ket{0}$ and $\ket{1}$ into their intermediate states $\ket{+}=\frac{1}{\sqrt{2}}(\ket{0}+\ket{1})$ and $\ket{-}=\frac{1}{\sqrt{2}}(\ket{0}-\ket{1})$ respectively, and it also satisfies $HH^\dagger=I$. (5)For any single-bit gate, it can be decomposed into two Z-Rotation gates and one Y-Rotation gate. i.e. Three parameters $(\phi,\lambda,\theta)$ can determine a unique unitary matrix. (6)The CNOT gate is composed of the control bit $c$ and the target bit $t$. When the control bit is $\ket{1}$, the target bit is reversed (using the Pauli-X gate). There may be more than one control bit called $C^n-NOT$ gate, which can entangle multiple qubits. It can be seen from \cite{nielsen2002quantum} that Y-Rotation gate, Z-Rotation gate, and CNOT gate can form a general basic gate library, i.e., any unitary transformation can be realized by using only these three gates.

\begin{table}[htbp!]
\setlength{\abovecaptionskip}{5pt}%
\setlength{\belowcaptionskip}{5pt}%
\caption{Common quantum gates.}
\label{s2t1}
\begin{center}
\begin{tabular}{c|c|c}
 \tabincell{c}{Pauli-X\\ gate}&  \tabincell{c}{Y-Rotation\\ gate}& \tabincell{c}{Z-Rotation\\ gate}\\
 \hline
$\Qcircuit @C=1em @R=.7em {
& \gate{X} & \qw
}$&$
\Qcircuit @C=1em @R=.7em {
& \gate{R_\mathrm{Y}(\theta)} & \qw
}$&$
\Qcircuit @C=1em @R=.7em {
& \gate{R_{Z_{\lambda}}} & \qw
}$\\
  \hline
$\begin{bmatrix} 0 & 1 \\ 1 & 0\end{bmatrix}$ & $\begin{bmatrix} \cos(\frac{\theta}{2}) &  \sin(\frac{\theta}{2}) \\  -\sin(\frac{\theta}{2}) &  \cos(\frac{\theta}{2})\end{bmatrix}$ &$\begin{bmatrix} 1 & 0 \\ 0 & e^{i\lambda}\end{bmatrix}$ \\
\Xhline{1.5pt}
\tabincell{c}{Hadamard\\ gate}&\tabincell{c}{Single-bit gate}&CNOT gate\\
 \hline
$
\Qcircuit @C=1em @R=.7em {
& \gate{H} & \qw
}$&$
\Qcircuit @C=1em @R=.7em {
& \gate{U} & \qw
}$&$
\Qcircuit @C=1em @R=.7em {
\lstick{c}& \ctrl{1} &  \qw \\
\lstick{t}& \targ &  \qw
}$\\
\hline

$\frac{1}{\sqrt{2}}\begin{bmatrix} 1 & 1 \\ 1 & -1\end{bmatrix}$
&
$\begin{bmatrix} \cos(\frac{\theta}{2}) & -e^{i\lambda}\sin(\frac{\theta}{2}) \\ e^{i\phi}\sin(\frac{\theta}{2}) & e^{i(\phi+\lambda)}\cos(\frac{\theta}{2})\end{bmatrix}$ &
$\begin{bmatrix} 1 & 0 & 0 & 0 \\ 0 & 1 & 0 & 0 \\ 0 & 0 & 0  &1 \\ 0&0 & 1&0 \end{bmatrix}$ \\
\end{tabular}
\end{center}
\end{table}

\subsubsection{computational advantages on quantum circuits}

The physical realization of quantum computers is still at an immature stage, and the advantages of quantum algorithms are mainly reflected in computational complexity. In quantum computing, a common time complexity calculation method is the number of basic gates (or the depth of the quantum circuit) required to implement the algorithm. According to the previous introduction, the matrix operation of the quantum gate can speed up the realization of the matrix multiplication operation. However, the addition also can be realized on quantum circuits, but it has no acceleration advantages because it is difficult to cope with the physical properties of superposition and entanglement.

In this section, we introduce the basic concepts of DST and quantum computing.

\section{BF-QC: Belief functions on quantum circuits}
\label{3}

In this section, we encode the BBA into the amplitudes of the quantum superposition state and propose its implementation.
\subsection{Basic Belief Assignment on quantum circuits}

For an $n$-element FoD $\Theta$ and its power set $2^{\theta}=\{F_{i}\}=\{\{\varnothing\},\{\theta_{1}\},\dots,\{\theta_{n}\},\{\theta_{1}\theta_{2}\},\dots,\{\theta_{1}\dots\theta_{n}\}\}$. According to the binary coding method proposed by Smets \cite{smets2002application}, a quantum superposition state under $n$ qubits can be expressed in Table \ref{s3t1}.

\begin{table*}[htbp!]
\caption{Each qubit corresponds to an element. When the quantum state is $\ket{1}$, the corresponding element exists in the focal element. When the qubit is $\ket{0}$, the intersection of the corresponding element and the focal element is empty.}
\label{s3t1}
\begin{center}
\begin{tabular}{c|c|c|c|c|c|c|c}
\hline
Quantum state &$\ket{0}^{\bigotimes n}$&$\ket{0}^{\bigotimes n-1}\ket{1}$&$ \ket{0}^{\bigotimes n-1}\ket{1}\ket{0}$ & $\dots$ & $\ket{0}^{\bigotimes n-2}\ket{1}^{\bigotimes 2}$& $\dots$ &$\ket{1}^{\bigotimes n}$\\
\hline
Focal set &$\{\varnothing\}$&$\{\theta_{1}\}$&$\{\theta_{2}\} $&$\dots$&$\{\theta_{1}\theta_{2}\}$&$\dots$&$\{\theta_{1}\dots\theta_{n}\}$\\
\hline
\end{tabular}
\end{center}
\end{table*}

\begin{definition}[BBA-QC]
\rm{
For a BBA $m$ under an $n$-element FoD, its encoding on the quantum state $\ket{m}$ is defined as follows:

\begin{equation}\label{qbpae}
\begin{aligned}
&\ket{{m}}=\sum_{j=0}^{2^{n-1}}\sqrt{m(F)}\ket{j}=\sqrt{m(\emptyset)}\ket{0\cdots 0}+\sqrt{m(\theta_{1})}\ket{0\cdots01}\\&+\sqrt{m(\theta_{2})}\ket{0\cdots10}+\dots+\sqrt{m(\theta_{1}\dots\theta_{n})}\ket{1\cdots1},
\end{aligned}
\end{equation}
where the amplitudes also can be a complex number, but they should satisfy that the square of them are equal to the mass functions of the corresponding focal sets.}
\end{definition}

In quantum states, the bases are mutually orthogonal, while in DST, the focal sets are not completely mutually exclusive. Hence, different from the definition in \cite{vourdas2014quantum,vourdas2014lower}, BBA-QC cannot achieve the physical coupling of quantum state and BBA, but this does not affect the advantages of using quantum computing to perform BBA operations. When a gate operates on a certain qubit, it can realize the same operation on the amplitudes of all quantum states whose corresponding bit is $1$. Similarly, in BBA-QC, when we operate on a certain element, we can change the mass functions of all focal sets which contain this element at the same time.

\subsection{Implement BBA on quantum circuits}

\begin{figure}[htbp!]
\includegraphics[width=0.7\textwidth]{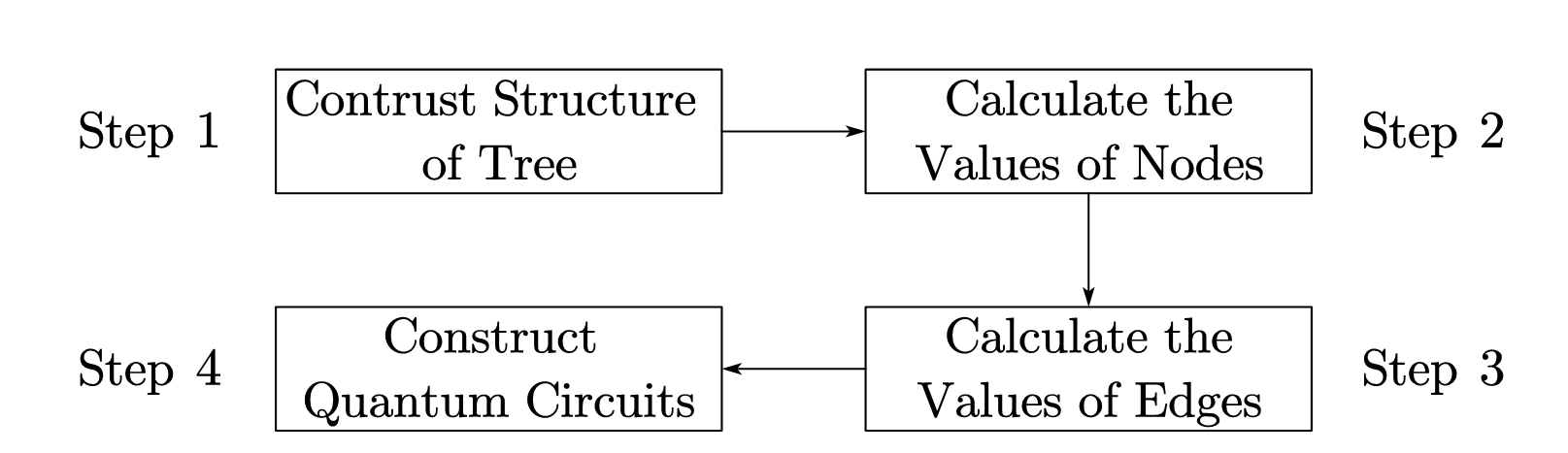}
\caption{The flow of implementation BBA-QC.}
\label{Q_implement_Flow}
\end{figure}

We propose a method for encoding classical BBA on quantum circuits based on the tree-like memory structure in \cite{prakash2014quantum}. Given a BBA under an $n$-element FoD, it can be implemented in time (depth) $\mathcal{O}(\log (2^n-1))$ by $\mathcal{O}(2^n-1)$ gates on quantum circuits. Figure \ref{Q_implement_Flow} shows the flow of implementation steps, and the specific operations are shown in Algorithm \ref{s3a1}

\begin{algorithm}[htbp!]
  \caption{Implement BBA-QC based on the tree-like memory structure.}
  \label{s3a1}
  \begin{algorithmic}[1]
    \Require
    A BBA $m$ under a $n$-element FoD;
    An initial quantum state $\ket{m_1}=\ket{1}^n$;
    \Ensure
     A quantum state of $m$: $\ket{m}=\sum_{i=0}^{2^n-1}\sqrt{m(F_i)}\ket{i}$
    \State \textbf{Construct the Structure of Tree:}
    \State \{
     \State         For the whole structure, from top to bottom, each layer operates a qubit, and the bit order is from left to right.
    \State For the each parent node $\ket{\Psi_\mathrm{parent}}$, $R_\mathrm{Y}(2\theta)\ket{1}\rightarrow\sin(\theta)\ket{0}+\cos(\theta)\ket{1}$, and rule the left child node is $\ket{0}$ and the right node is $\ket{1}$.
    \State \}
    \State \textbf{Calculate the Values of Nodes:}
    \State \{
    \State In $n$th layer, the nodes' values are $v^n_i=\sqrt{m(F_i)}$.
    \State \For {Layer $l$ form $n-1$ to $1$}
    \State $v^{l}=\sqrt{(v^{l+1}_\mathrm{left\_child\_node})+(v^{l+1}_\mathrm{right\_child\_node})}$\\ \% //There are $2^{l-1}$ nodes in $l$th layer.//
    \EndFor
    \State \}
    \State \textbf{Calculate the Values of Edges:}
    \State \{
    \State \For {Each parents node $\ket{\Psi_\mathrm{parent\_j}}$}
    \State $\theta_j=\arctan(\sqrt{\frac{v_\mathrm{left\_child\_node}}{v_\mathrm{right\_child\_node}}})$ \% //$v_\mathrm{child\_node}$ is value of child node of $\ket{\Psi_\mathrm{parent\_j}}$//
    \State The value of left edge is $\sin(\theta_j)$ and the right is $\cos(\theta_j)$
    \EndFor
    \State \}
    \State \textbf{Construct Quantum Circuits:}
    \State \{
    \State \For{Each $\theta_j$ of edge}
    \State $R_\mathrm{Y}(2\theta_j)=\begin{bmatrix} \cos(\theta_j) &  \sin(\theta_j) \\  -\sin(\theta_j) &  \cos(\theta_j)\end{bmatrix}$
    \EndFor
    \State \For{Layer $l$ from $1$ to $n$}
    \State Operate $C^{l-1}-R_\mathrm{Y}(2\theta)$ on $\ket{m_{l-1}}$ to generate $\ket{m_{l}}$.
    \EndFor
    \State $\ket{m}=\ket{m_{n}}$
    \State \}
    \State\Return $\ket{{m}}$;
  \end{algorithmic}
\end{algorithm}

\begin{figure}[htbp!]
\centering
\includegraphics[width=0.99\textwidth]{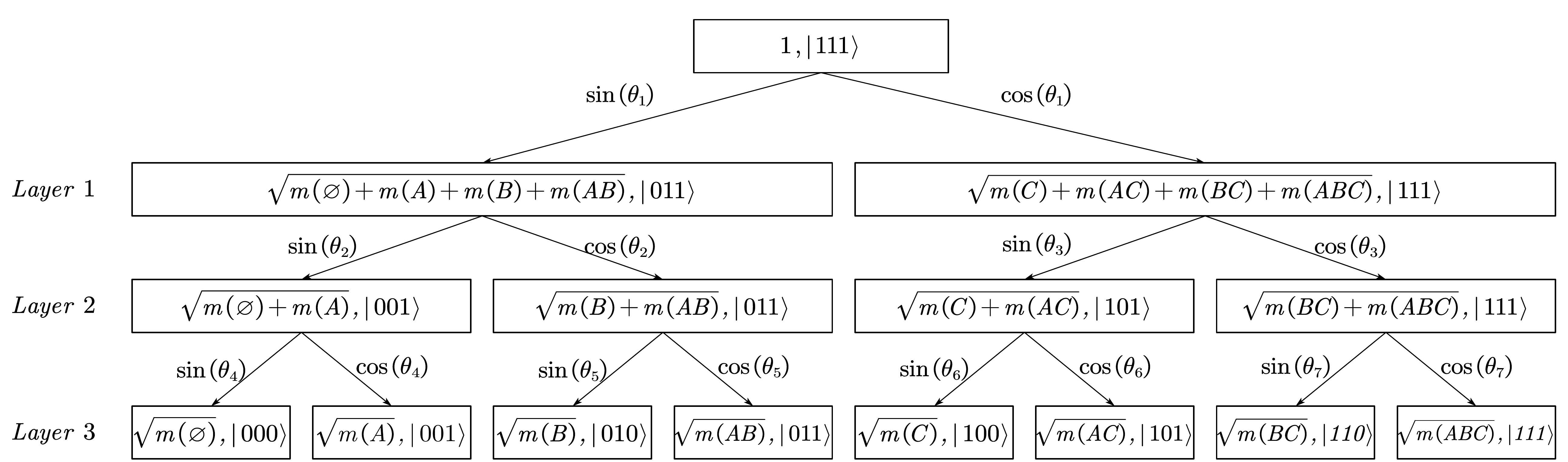}
\caption{The structure is similar to the hierarchical hypothesis space \cite{Jean1985hierarchical}, only one qubit per operation. }
\label{Q_Tree}
\end{figure}

\begin{figure}[htbp!]
\centering
\footnotesize
\Qcircuit @C=2.5em @R=1em {
&\mbox{$1^{st} Layer$} &  \mbox{$2^{nd} Layer$}& & \mbox{$3^{rd} Layer$}& & & & &\\
\lstick{\ket{1}} &\gate{R_{Y}(2\theta_{1})}& \ctrlo{1}                             & \ctrl{1}                             &\ctrlo{1} &\ctrlo{1}&\ctrl{1}&\ctrl{1}&\qw&\rstick{}\qw\\
\lstick{\ket{1}} &\qw                                  & \gate{R_{Y}(2\theta_{2})} & \gate{R_{Y}(2\theta_{3})}&\ctrlo{1}&\ctrl{1}&\ctrlo{1}&\ctrl{1}&\qw&\rstick{}\qw\\
\lstick{\ket{1}} &\qw                                  & \qw                                   & \qw                                  &\gate{R_{Y}(2\theta_{4})}&\gate{R_{Y}(2\theta_{5})}&\gate{R_{Y}(2\theta_{6})}&\gate{R_{Y}(2\theta_{7})}&\qw&\rstick{}\qw
\gategroup{2}{2}{2}{2}{.7em}{--}\gategroup{2}{3}{3}{4}{.7em}{--}\gategroup{2}{5}{4}{8}{.7em}{--}
}
\caption{According to the structure in Figure \ref{Q_Tree}, the controlled Y-Rotation gates can be used to simulate the evolution of each layer. To popularize this circuit, for $n$-element FoD, BBA-QC can be implemented with $n$ qubits and $2^{n}-1$ quantum gates, and for the $k$th layer, $2^{k-1}$ gates are needed to simulate.}
\label{s3f2}
\end{figure}
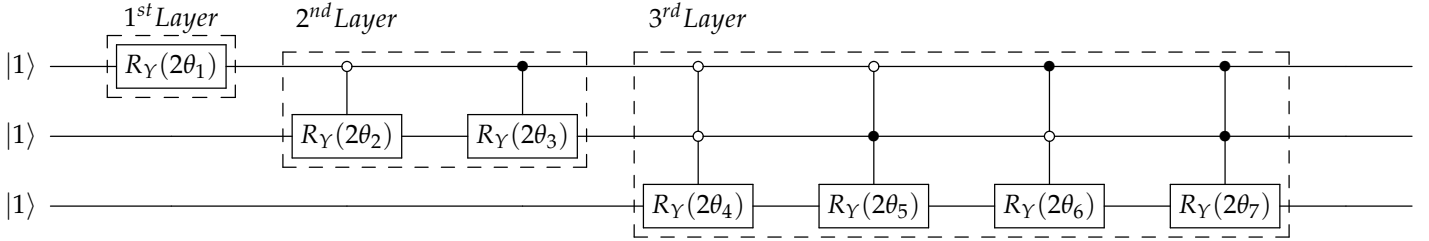

In order to show the process intuitively, the tree-like structure memory and implementation circuits of the BBA under $3$-element FoD $\{A, B, C\}$ are shown in Figure \ref{Q_Tree} and \ref{s3f2}. When there are enough qubits, the depth of the implementation quantum circuit can be reduced to $\mathcal{O}(n)$. Precise implementation is usually an expensive cost in quantum computing, but our method has a relatively high application value due to the connection of power sets between DST and quantum computing.

In addition, QRAM \cite{giovannetti2008quantum} is also a fast way to extract quantum information. Compared with classical RAM, QRAM can store the classical information and extract them to quantum information. The architecture of QRAM is shown in \cite{giovannetti2008architectures}, which realizes an exponential acceleration than classical queries and leads that quantum algorithms still be superior to classical algorithms even if they may need constant implementations and measurements. Therefore, in this paper, we suppose that BBA-QC has been extracted by QRAM to produce a quantum state.

In this section, we define the form of BBA on quantum circuits, and give the corresponding implementation method, which theoretically realizes the encoding of BBA into a quantum state. Hence, one pure quantum state can carry the information of a BBA.

\subsection{Extract belief functions from BBA-QC} \label{QBF}

According to Equation \ref{bele} - \ref{qe}, the belief functions express the total belief of focal sets with common elements. However, performing a traversal search to calculate the belief functions has high complexity. Smets simplify the process by matrix operations. For $Bel$, $Pl$, and $q$ function, it can be generated by reversible matrices $\boldsymbol{M_{Bel}}$, $\boldsymbol{M_{Pl}}$ and $\boldsymbol{M_{q}}$. Suppose the focal sets $F$ and $G$ represent the row and column of matrix $\boldsymbol{M}$,

\begin{equation}\label{belief_m}
M_{Bel}(F,G)=
\begin{cases}
1 & \text{iff $G\subseteq F$}\\
0 & \text{Others}
\end{cases};
M_{q}(F,G)=
\begin{cases}
1 & \text{iff $F\subseteq G$}\\
0 & \text{Others}
\end{cases};
M_{Pl}(F,G)=
\begin{cases}
1 & \text{iff $F\cap G\neq\varnothing$}\\
0 & \text{Others}
\end{cases}.
\end{equation}
Given a BBA $m$ under $n$-element FoD, the complexity of $\boldsymbol{M_{Bel}}\cdot\boldsymbol{m'}$ and  $\boldsymbol{M_{q}}\cdot\boldsymbol{m'}$ both are $\mathcal{O}(\frac{1}{2}4^n)$, and the complexity of $\boldsymbol{M_{Pl}}\cdot\boldsymbol{m'}$ is $\mathcal{O}(4^n)$. Since the range of belief functions are $[0, 1]$, when belief functions are encoded into the quantum states, a quantum bit carry the information of a belief function of a focus set.

\begin{definition}[BF-QC]\label{qbelf} \rm{
Given a BBA $m$ under an $n$-element FoD $\Theta$, the quantum states of Bel functions, Pl functions and q functions of focal sets $F$ are denoted as
\begin{equation}\label{qbelfe1}
\begin{aligned}
&\ket{Bel(F)}=\sqrt{Bel(F)}\ket{1}+\sqrt{Pl(\overline{F_{i}})}\ket{0};\\
&\ket{{Pl(F)}}=\sqrt{Pl(F)}\ket{1}+\sqrt{Bel(\overline{F})}\ket{0};\\
&\ket{{q(F)}}=\sqrt{q(F)}\ket{1}+\sqrt{1-q(F)}\ket{0}.
\end{aligned}
\end{equation}}
\end{definition}

\begin{figure*}[htbp!]
\centering
\includegraphics[width=0.98\textwidth]{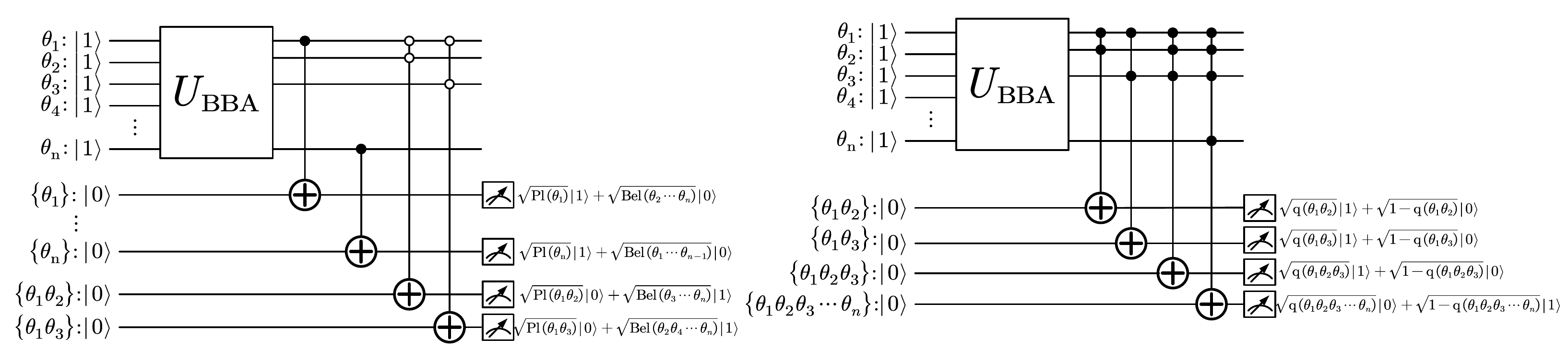}
\caption{$U_\mathrm{BBA}$ is the implementation process of $\ket{m}$. The black control bit indicates the execution target qubit when the control qubit is $1$, and the white control bit indicates the execution target qubit when the control qubit is $0$. Due to the relationship between the Bel function and the $Pl$ function, when a $C^n$-NOT gate is executed, a quantum state can be implemented, which contains both the focal set's Bel function and its complement's $Pl$ function.}
\label{BF-QC}
\end{figure*}

In quantum computing, entanglement between qubits is achieved through control gates. Similarly, we extract the required belief from the BBA by the C-NOT gates. Selecting the appropriate control bits can encode the belief functions into the amplitude of the target qubit. After constant numbers measurements, the classical belief functions can be observed. Figure \ref{BF-QC} shows circuits for the implementation of $Bel$ function, $Pl$ function, and $q$ function of $m$ under FoD $\Theta$. Although capturing classical belief functions needs to $t$ times implementations and measurements\footnote{$\Qcircuit @C=1em @R=.7em {& \meter & \qw }$ means measurements.}, it is only necessary to re-prepare the BF-QC rather than BBA-QC. One control qubit can extract the belief function of a focal set, so the $\mathrm{O}(2^n)$ time complexity is required to implement all belief functions of BBA. Compared with the classical matrix calculation, an exponential speedup is achieved.

After BF-QC implemented from BBA-QC, it is not the only purpose to extract it as classical data by measurements. Although the quantum data processing of BF-QC is not the focus in this paper, continuing to process them with quantum algorithms may also be an option worth discussing.

\subsection{Simulate the BF-QC on Qiskit}

In order to prove the rationality of the proposed model, we simulated the implementation process of a BBA $m$ under $3$-element FoD on the Qiskit platform (IBM (2016). IBM quantum experience. URL:https://quantum-computing.ibm.com) in Example \ref{s3e3}.

\begin{example}\label{s3e3}\rm{
Given a BBA $m=\{m(A)=\frac{1}{18}, m(B)=\frac{1}{6},m(C)=\frac{1}{6},m(AB)=\frac{1}{9},m(AC)=\frac{1}{18},m(BC)=\frac{2}{9},m(ABC)=\frac{2}{9}\}$ under the FoD $X=\{A,B,C\}$. According to Equation \ref{ple} and \ref{qe}, $Pl(C)=\frac{2}{3}$ and $q(BC)=\frac{4}{9}$. According to the Algorithm \ref{s3a1} and Figure \ref{Q_Tree}, we construct a quantum circuit on Qiskit to implement the $\ket{{m}}$ in Figure \ref{s3f7}, and the simulation result with $1024$ times measurements are shown in Figure \ref{s3f8}. In addition, $\ket{{Pl(C)}}$ and $\ket{{q(BC)}}$ are also simulated based on Figure \ref{BF-QC}, and Figure \ref{s3f9} and \ref{s3f10} show their probability amplitude after measurements.

\begin{figure*}[htbp!]
\centering
\includegraphics[width=0.95\textwidth]{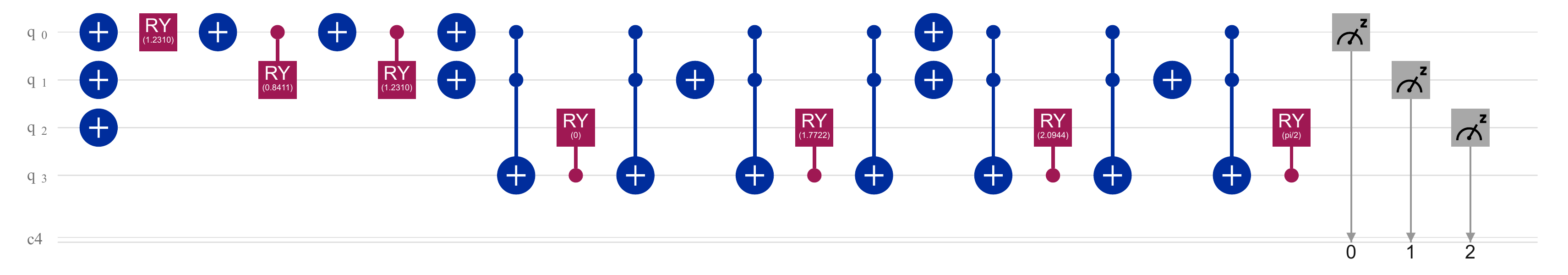}
\caption{Based on the QRAM, use NOT gate, Y-Rotation gate, CNOT gate and CCNOT gate to implement the QBBA $\ket{\Psi_{m(X)}}$.}
\label{s3f7}
\end{figure*}

  \begin{figure}
  \begin{minipage}[htbp]{0.5\linewidth}
    \centering
    \includegraphics[width=0.98\textwidth]{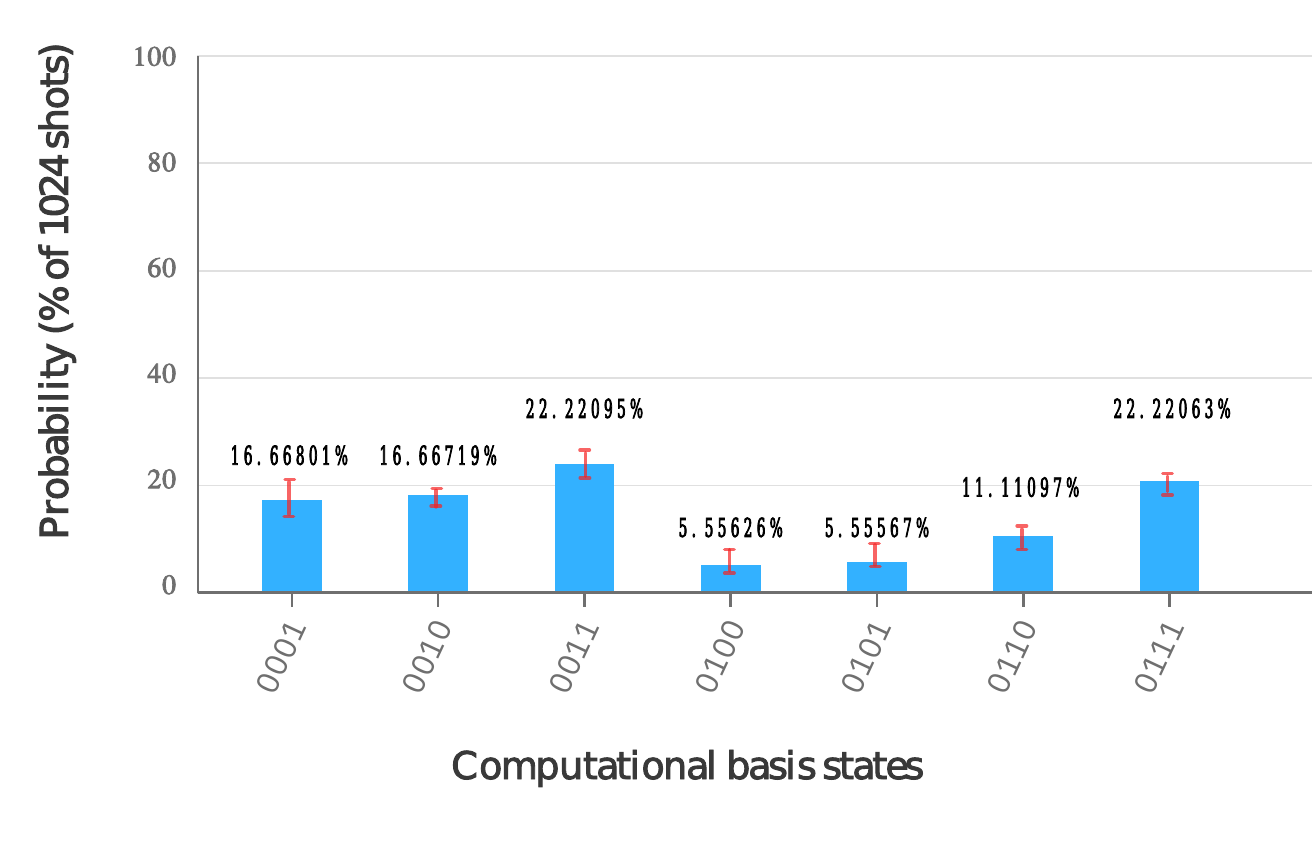}
    \caption{BBA is implemented on the amplitude of corresponding state.}
    \label{s3f8}
  \end{minipage}
  \begin{minipage}[htbp]{0.24\linewidth}
    \centering
    \includegraphics[width=0.98\textwidth]{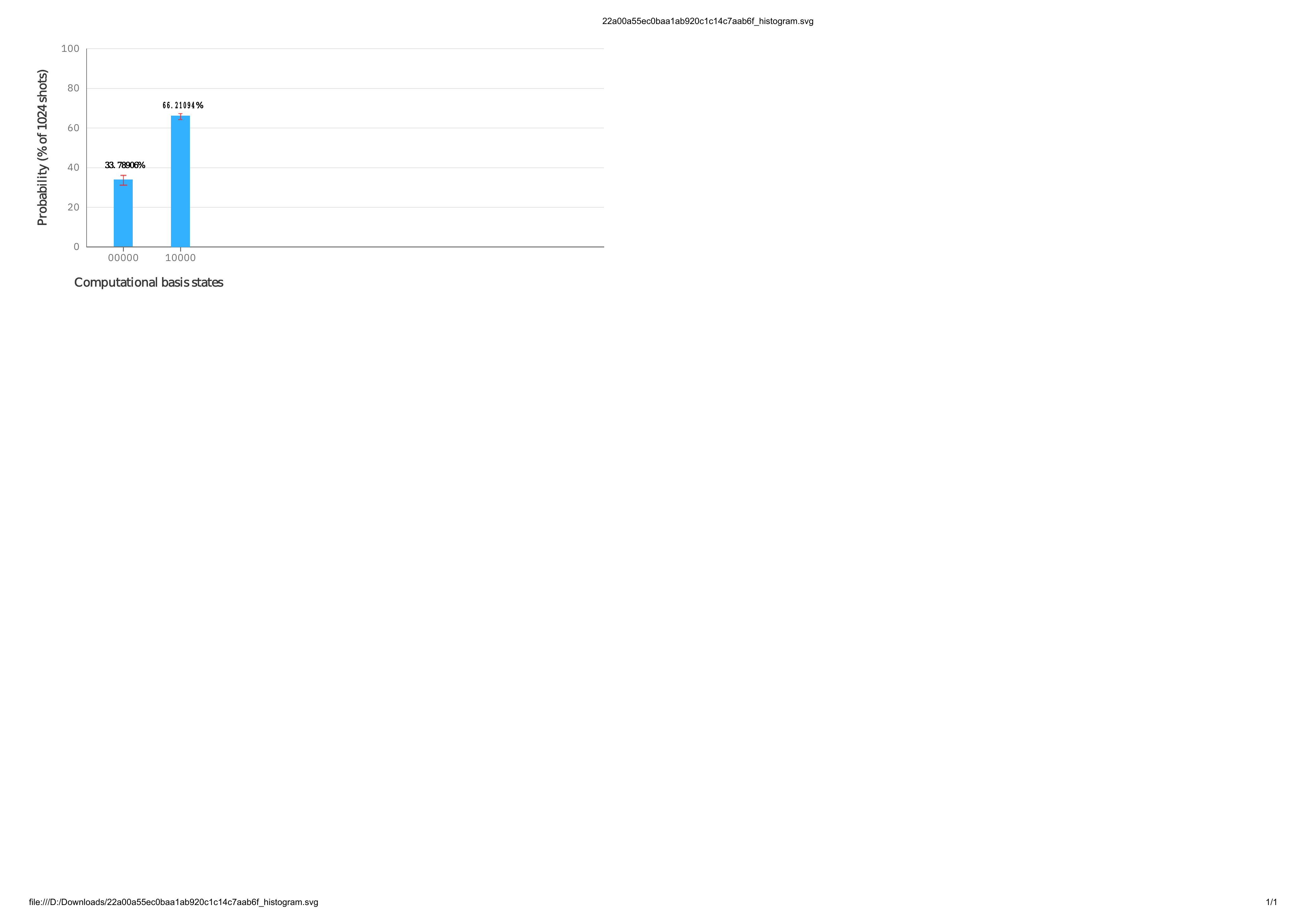}
    \caption{When the first bit is $\ket{1}$, the amplitude of the state is belong to $Pl(C)$.}
    \label{s3f9}
  \end{minipage}
 \begin{minipage}[htbp]{0.24\linewidth}
    \centering
    \includegraphics[width=0.98\textwidth]{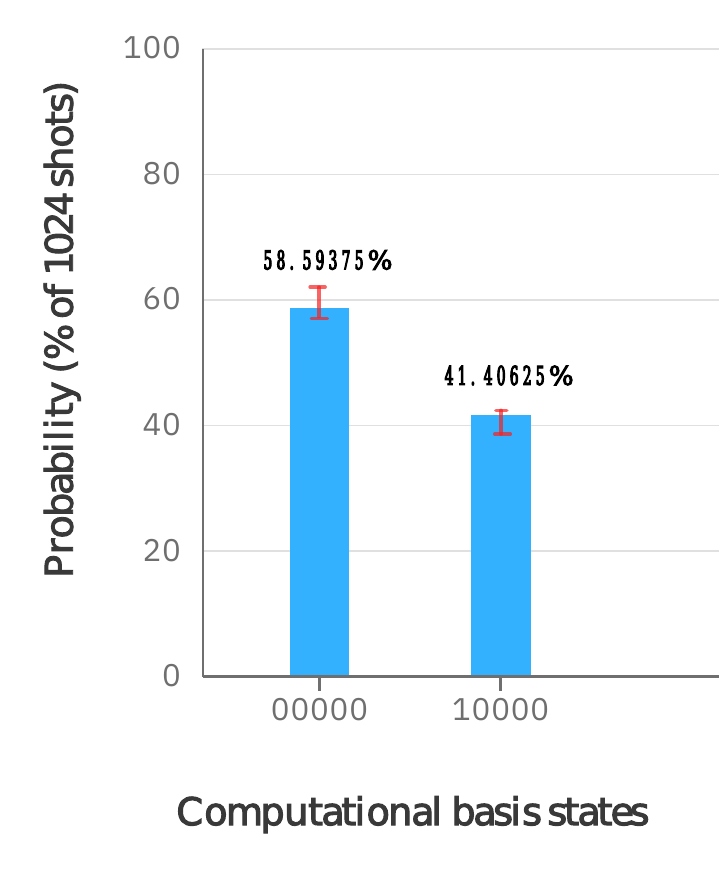}
    \caption{The probability amplitude of  a $q(BC)$,}
    \label{s3f10}
  \end{minipage}
\end{figure}}
\end{example}

Based on the above results, the errors of measured results are within $10^{-3}$, and these errors come from the quantum operations themselves. It is proved that the quantum state can store the classical BBA information, and in the extraction of the belief functions information, it achieves an exponential speedup compared with the classical matrix operation.

\section{Operate belief functions on quantum circuits}
\label{4}
We have previously achieved the implementation of the basic information unit (BBA and belief functions) of DST. According to Section \ref{pre}, common algorithms in DST can be represented as matrices operations. However, quantum computing only has advantages in unitary matrix operations. In order to make the matrix operations of DST perform on quantum circuits, we first give a matrix operation on the quantum circuit based on the HHL algorithm, which can relax the requirements to invertible matrix.

\subsection{Matrix operation of BBA on quantum circuits}

HHL algorithm is proposed in 2008. Under the premise of using QRAM to extract quantum states, it can exponentially accelerate the solution of linear system problems (LSP) on quantum circuits. When $\boldsymbol{A}$ is a Hermitian matrix and $\boldsymbol{b}$ is a unit vector, HHL converts $\boldsymbol{A}\boldsymbol{x}=\boldsymbol{b}$ into $\boldsymbol{x}=\boldsymbol{A^{-1}}\boldsymbol{b}$ and solve, and obtain $\ket{\widetilde{x}}$ containing the information of $\boldsymbol{x}$. Based on above, we applied HHL algorithm to operate BBAs on quantum circuits using invertible matrices.

\subsubsection{problem description}
Given a evolution matrix $\boldsymbol{M}$ and a BBA-QC $\ket{m}$ under an $n$-element FoD. Target BBA $m_0$ meets $\boldsymbol{m_0'}=\boldsymbol{M}\cdot\boldsymbol{m'}$. Since the initial $\boldsymbol{m'}\rightarrow\ket{m}$, matrix $\boldsymbol{M}$ is changed to $\boldsymbol{M}_\mathrm{N}=\boldsymbol{M}\cdot diag(\sqrt{\boldsymbol{m}})$. The whole process is written as $\boldsymbol{M}_\mathrm{N}\ket{m}\rightarrow\ket{m_0}$. The ideal output state is the normalized of $\boldsymbol{m_0'}$, $\ket{{m_0}}=\frac{\sum_{i=0}^{2^n-1}m_0(F_i)\ket{i}}{||\sum_{i=0}^{2^n-1}m_0(F_i)\ket{i}||_2}$. The output state $\ket{\widetilde{m_0}}$ satisfying $||\ket{\widetilde{m_0}}-\ket{{m_0}}||_2\leq\epsilon$ can be measured with probability larger $\frac{1}{2}$. Due to the normalization limitation of the quantum state amplitude, for normal matrices (not satisfying $\boldsymbol{M}\boldsymbol{M}^{\dagger}\neq \boldsymbol{I}$), the output state can only contain partial information of $\boldsymbol{m_0}$. Because of the limitation of unitary operations, $\boldsymbol{M}_\mathrm{N}$ should be an invertible matrix. If $\boldsymbol{M}_\mathrm{N}$ is Hermitian, it can be directly executed evolution, otherwise it should be written as $\begin{bmatrix} 0 & \boldsymbol{M}_\mathrm{N}^{\dagger} \\ \boldsymbol{M}_\mathrm{N} & 0\end{bmatrix}$, and the corresponding quantum state is written as $\begin{bmatrix} \ket{m} \\ \boldsymbol{0}\end{bmatrix}$.

\subsubsection{matrix evolution BBA-QC}

\begin{figure}[htbp!]
\centering
\includegraphics[width=0.75\textwidth]{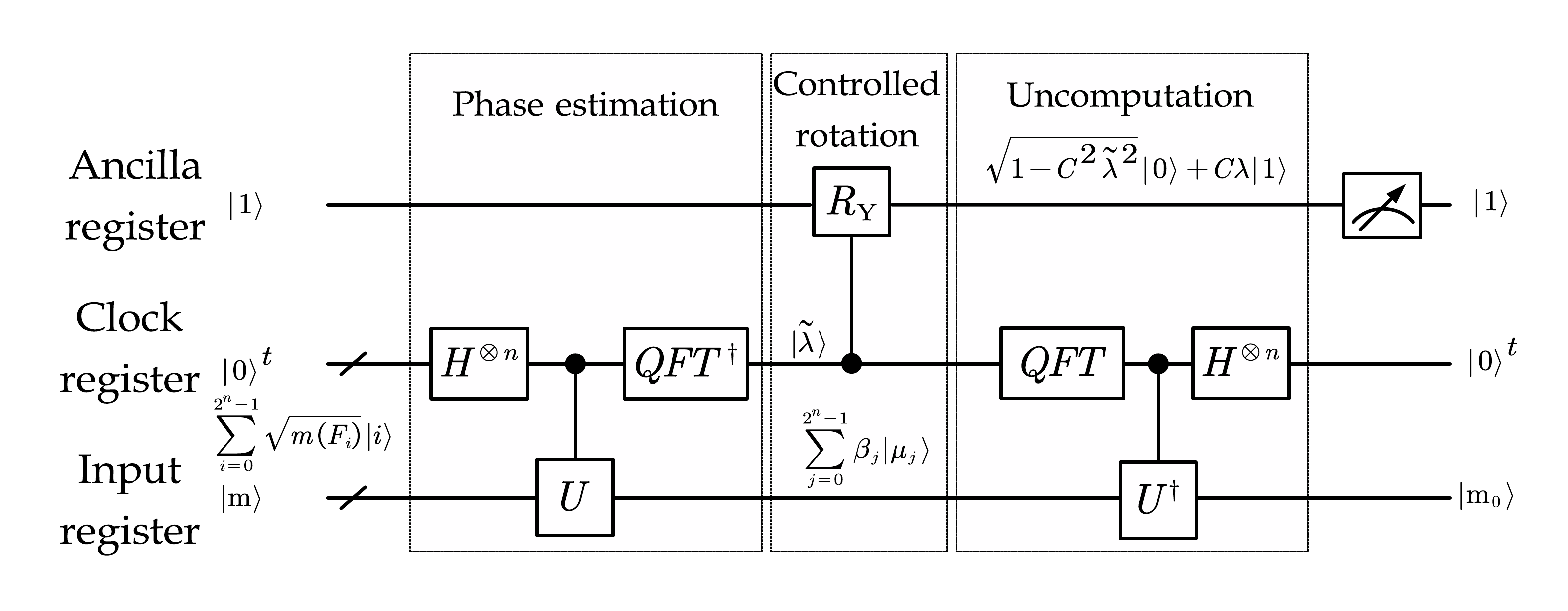}
\caption{The quantum circuits of MEoB.}
\label{MEoB}
\end{figure}

The process of algorithm is divided into three steps, which are phase estimation, controlled rotation and uncomputation. Phase estimation, as a commonly used in quantum algorithm, extract the phase of the unitary matrix with two quantum registers. In matrix evolution of BBA-QC (MEoB), since $\boldsymbol{M}_\mathrm{N}$ is Hermitian, $\exp(i\boldsymbol{M}_\mathrm{N}t)$ is a unitary matrix and satisfies $\boldsymbol{U}\ket{\mu_j}=\exp(i\boldsymbol{M}_\mathrm{N}t)\ket{\mu_j}=\exp(i\lambda_j t)\ket{\mu_j}$, where $\lambda_j$ and $\ket{\mu_j}$ are eigenvalue and eigenvector of $\boldsymbol{M}_\mathrm{N}$. By estimating the phase of the unitary matrix $\boldsymbol{U}=\exp(i\boldsymbol{M}_\mathrm{N}t)$, the eigenvalues $\lambda_j$ can be extracted. The second step is controlled rotation. Through C-RY gates, the eigenvalue extracted in the first step are rotated into the amplitude of $\ket{m}$ to complete the matrix operation. The third step is to undo previous calculations, i.e., use Hermitian conjugation to reverse the calculation of the first step. After the above operations, $\ket{m}$ is evolved to $\ket{m_0}$. The simplified circuits of MEoB are shown in Figure \ref{MEoB}, which is similar with HHL algorithm \cite{DUAN2020hhl}.

\textbf{Step 1: phase estimation:}

\begin{figure*}[htbp!]
\centering
\includegraphics[width=0.9\textwidth]{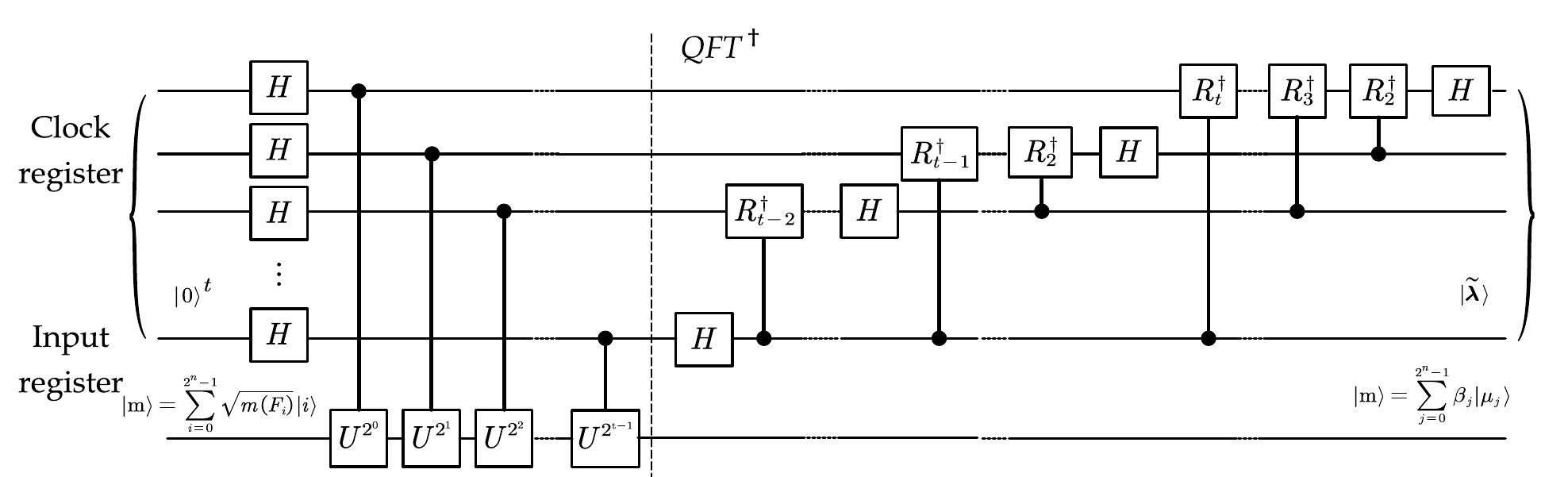}
\caption{The clock register is to simulate the evolution of time, apply the desired phase to the superposition state, and then use the inverse quantum Fourier transform (the specific process of quantum Fourier transform can refer to \cite{nielsen2002quantum}) to obtain a binary representation of the phase with a precision of $t$. Input register is entangled with clock register by C-U gates, where $U=\exp(i M_\mathrm{N}t)$, and the output is transformed the basis to eigenvalues of $M_\mathrm{N}$.}
\label{QPE}
\end{figure*}

The process of phase estimation is shown in Figure \ref{QPE}. First, initial state is performed $t$ Hadamard gates $H^{\bigotimes t}$
\begin{equation}
    \ket{0}^t\ket{m}\stackrel{H^{\bigotimes t}\bigotimes I}{\longrightarrow}\left(\frac{1}{\sqrt{2}}\right)^t(\ket{0}+\ket{1})^{t}\ket{m}.
\end{equation}
For the quantum circuit \includegraphics[width=0.14\textwidth]{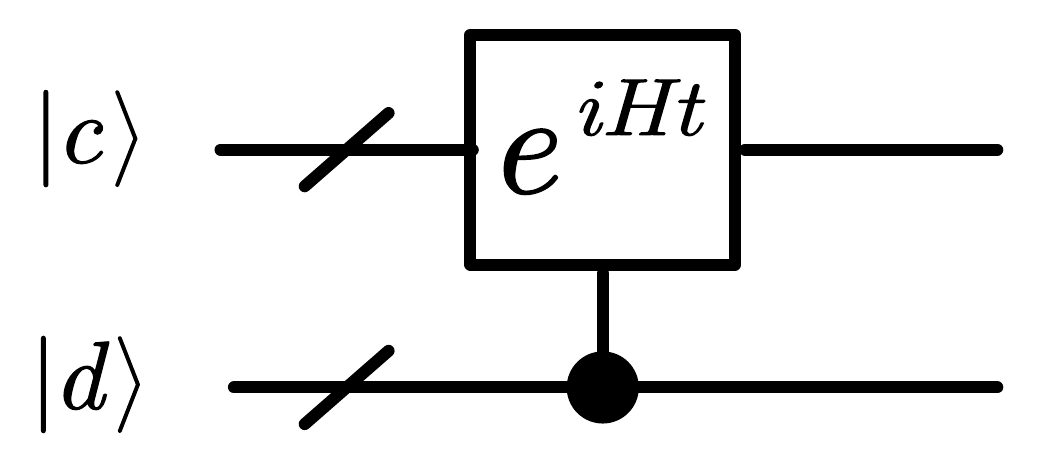}, it can be formed as $\ket{c}\ket{d}\rightarrow \sum_{i}\sum_{j}\exp(iHtd_{j})\ket{c_i}\ket{d_j}$. When the $C-U$ gates are performed, the state of clock register is
\begin{equation}
\begin{aligned}
      &\left(\frac{1}{\sqrt{2}}\right)^t(\ket{0}+\ket{1})^{t}\stackrel{t\cdot \text{C-U}}{\longrightarrow}  \left(\frac{1}{\sqrt{2}}\right)^t\left(\ket{0}+\exp(i2^{t-1}\lambda )\ket{1}\right)\\
      &\cdots\left(\ket{0}+\exp(i2^{0}\lambda )\ket{1}\right)=\left(\frac{1}{\sqrt{2}}\right)^t\sum_{k=0}^{2^t-1}\exp(i\lambda k)\ket{k}.
\end{aligned}
\end{equation}
This step is to put the target eigenvalue $\lambda$ as the phase in the quantum state, so that the phase can be shown in binary form with $t$ precision using the inverse quantum Fourier transform ($\mathrm{QFT}^\dagger$)

\begin{equation}
\begin{aligned}
    \left(\frac{1}{\sqrt{2}}\right)^t\sum_{k=0}^{2^t-1}\exp(i\lambda k)\ket{k}\stackrel{\mathrm{QFT}^\dagger}{\longrightarrow}\left(\frac{1}{2}\right)^t \sum_{x=0}^{2^t-1}\sum_{k=0}^{2^t-1}\exp(\frac{2\pi i}{2^t}-\lambda i)\ket{x},
\end{aligned}
\end{equation}
where $\widetilde{\lambda}=\lambda \approx -\frac{2\pi x}{2^t}$. Since the quantum state can be decomposed under any standard orthonormal basis, no operation needs to be performed on the input state. When the basis are the eigenvectors of $\boldsymbol{M}_\mathrm{N}$, it is written as $\ket{m}=\sum_{i=0}^{2^n-1}\sqrt{m(F_i)}\ket{i}\rightarrow\sum_{j=0}^{2^n-1}\ket{\mu_j}\braket{\mu_j|m}=\sum_{j=0}^{2^n-1}\beta_j\ket{\mu_j}$. Based on above, phase estimation realized the following operation
\begin{equation}
    \ket{0}\ket{m}=\sum_{j=0}^{2^n-1}\beta_j\ket{0}\ket{\mu_j}\stackrel{\mathrm{PE}}{\longrightarrow}\sum_{j}^{2^n-1}\sum_{k}^{2^t-1}\alpha_{j,k}\beta_k\ket{\widetilde{\lambda}_k}\ket{\mu_j},
\end{equation}
where $\alpha_{j,k}$ is a constant generated during operation, and it equals $1$ in the ideal situation. In the following derivation, we assume that $\alpha_{j,k}=1$.

\textbf{Step 2: controlled rotation}

The purpose of this step is to operate the input state with the eigenvalues extracted in the previous step through the $C-RY$ gates, which realizes the matrix evolution. A third register (ancilla register) is added to complete the rotation operation,
\begin{equation}
\begin{aligned}
 \sum_{j=0}^{2^n-1}\beta_j\sum_{k=0}^{2^t-1} \ket{1}\ket{\widetilde{\lambda}_k}\ket{\mu_j}  \stackrel{C-RY(\theta)}{\longrightarrow} \sum_{j=0}^{2^n-1}\beta_j\sum_{k=0}^{2^t-1}\left (\sqrt{1-C^2\widetilde{\lambda}_k^2}\ket{0}+C\widetilde{\lambda}_k\ket{1}\right )\ket{\widetilde{\lambda}_k}\ket{\mu_j},
\end{aligned}
\end{equation}
where $\theta$ of C-RY gates is $\theta=\arccos(C\widetilde{\lambda}_k)$. After the rotation, if the measurements of ancilla register is $\ket{1}$, the goal eigenvalue has been added to the amplitude of the $\ket{\mu_j}$.

\textbf{Step 3: uncomputation}

 This step is Hermitian conjugate of the phase estimation, using the reversible feature of quantum computing to return the eigenvalue $\widetilde{\lambda}$ into the phase to ensure that there is no excess amplitude. The process of Dirac notations is
\begin{equation}
\begin{aligned}
    &\sum_{j=0}^{2^n-1}\beta_j\sum_{k=0}^{2^t-1}\left (\sqrt{1-C^2\widetilde{\lambda}_k^2}\ket{0}+C\widetilde{\lambda}_k\ket{1}\right )\ket{\widetilde{\lambda}_k}\ket{\mu_j}\stackrel{\mathrm{PE}^\dagger}{\longrightarrow}  \\&\sum_{j=0}^{2^n-1}\beta_j\sum_{k=0}^{2^t-1}\left (\sqrt{1-C^2\widetilde{\lambda}_k^2}\ket{0}+C\widetilde{\lambda}_k\ket{1}\right )\ket{0}^t\ket{\mu_j}.
\end{aligned}
\end{equation}

And then we can measure the ancilla register, and when the state is $\ket{1}$, i.e. project the state on $\ket{1}$-space, the input register is $\ket{m_0}$

\begin{equation}
\begin{aligned}
&\sum_{j=0}^{2^n-1}\beta_j\sum_{k=0}^{2^t-1}\left (\sqrt{1-C^2\widetilde{\lambda}_k^2}\ket{0}+C\widetilde{\lambda}_k\ket{1}\right )\ket{0}^t\ket{\mu_j}\stackrel{\ket{1}\bra{1}}{\longrightarrow}\sum_{j=0}^{2^n-1}\beta_j C\widetilde{\lambda}_j\ket{1}\ket{0}^t\ket{\mu_j}
\end{aligned}
\end{equation}

\begin{equation}
    \ket{m_0}=\frac{1}{\sqrt{\sum_{j=0}^{2^n-1}\beta_j^2C^2\widetilde{\lambda}^2}}\sum_{j}^{2^n-1}\beta_j C\widetilde{\lambda}_j\ket{\mu_j}=\frac{\boldsymbol{m_0'}}{||\boldsymbol{m_0'}||_2}.
\end{equation}

The core idea of MEoB is to extract the eigenvalues of $M_\mathrm{N}$, and put the eigenvalues on $\ket{m}$ under the eigenvector $\ket{\mu_j}$ basis.Based on the MEoB, all belief function operations implemented by invertible matrices can be performed on quantum circuits.

\subsubsection{complexity Analysis}
Reasonable analysis the complexity of MEoB is the key to find whether quantum algorithms have the advantage of handling BBAs. Suppose we have implemented the BBA-QC, the main complexity sources are phase estimation and controlled rotation. For phase estimation, it has been proven to out put the goal state in time $\mathcal{O}(X_U\log(2^n)/\epsilon)$, where $X_U$ is the time to simulate the Hamiltonian and $\epsilon$ is the error $|\lambda_j-\widetilde{\lambda}_j|\leq \epsilon$. For the controlled rotation, the complexity is $\mathcal{O}(\kappa)$, where $\kappa=\frac{|\lambda_{\max}|}{|\lambda_{\min}|}$ is the conditional numbers of matrix. In \cite{low2019hamiltonian}, it proved that the $s$-sparse \footnote{$s$ represents the maximum non-zero elements numbers in a row.} Hamiltonian can be simulated within $\mathcal{O}(s+\frac{\log (1/\epsilon)}{\log\log(1/\epsilon)})$. Hence the maximum complexity of MEoB is $\mathcal{O}((2^n+\frac{\log (1/\epsilon)}{\log\log(1/\epsilon)})n/\epsilon+\kappa)$,  compared with the classical complexity $\mathcal(4^n)$, BBA operations realize acceleration on quantum circuits.

\subsection{Similarity of BBA-QC}

In Section \ref{dbba}, since $d_{BBA}\in[0,1]$, we take $1-d_{BBA}$ as a standard of similarity measure, and propose BBA-QC similarity measure based on the fidelity.

\subsubsection{Fidelity and swap tests}

Fidelity is a widely accepted similarity measure of quantum states. When measuring the similarity of mixed states, the fidelity is a function about density matrix. However, when measuring pure states, fidelity can be written as the inner product of unit vectors. For $2$ probability distributions $\boldsymbol{p_1}$ and $\boldsymbol{p_2}$ under $N$-dimensional random variable, they can be represented as $\ket{ p_i}=\sum_{j=1}^N\sqrt{p_i(j)}\ket{j}$ and the fidelity is $Fid(\boldsymbol{p_1},\boldsymbol{p_2})=\sum_{j=1}^N\sqrt{p_1(j)p_2(j)}$. In classical computation the complexity of fidelity is $\mathcal{O}(3N)$, and Buhrman \textit{et al.} \cite{buhrman2001quantum} implemented it on quantum circuits called swap test with $\mathcal{O}(\log N)$.

Besides the goal states $\ket{p_1}$ and $\ket{p_2}$, an ancilla register $\ket{0}$ is added to store the fidelity of them. The Dirac notations of swap test is

\begin{equation}
\begin{aligned}
    &\ket{0}\ket{p_1}\ket{p_2}\stackrel{H\bigotimes I^{\bigotimes 2N}}{\longrightarrow}\frac{1}{\sqrt{2}}(\ket{0}+\ket{1})\ket{p_1}\ket{p_2}\stackrel{C-SWAP}{\longrightarrow}\\
    &\frac{1}{\sqrt{2}}\ket{0}\ket{p_1}\ket{p_2}+\frac{1}{\sqrt{2}}\ket{0}\ket{p_1}\ket{p_2}\stackrel{H\bigotimes I^{\bigotimes 2N}}{\longrightarrow}\frac{1}{\sqrt{2}}\ket{0}(\ket{p_1}\ket{p_2}+\\
    &\ket{p_2}\ket{p_1})+\frac{1}{\sqrt{2}}\ket{1}(\ket{p_1}\ket{p_2}-\ket{p_2}\ket{p_1})\stackrel{Pr(\ket{1})}{\longrightarrow}\frac{1}{2}+\frac{1}{2}|\braket{p_1|p_2}|^2,
\end{aligned}
\end{equation}
where C-SWAP represents when the control qubit is $\ket{1}$, swap two target qubits, and $Pr(\ket{1})$ represents measure the probability amplitude of $\ket{1}$. Hence, we can extract the fidelity of $\ket{p_1}$ and $\ket{p_2}$ by measuring constant times ancilla register.

\subsubsection{Fractal-based Belief (FB) inner product}

In DST, the fidelity formula first exploited in \cite{RISTIC2006276}. Because the focal sets are not mutually exclusive, it is unreasonable to perform the swap test operation directly on the BBA-QC. Although Jousselme \textit{et al.} gives many belief information measures in \cite{jousselme2012distances}, they have the following two problems: (1) There is addition or subtraction of BBAs, which has no complexity advantage in quantum circuits (2) The operation of putting a row vector on a matrix cannot be implement directly on a quantum circuit. Therefore, we use Fractal-based Belief (FB) inner product to measure similarity of BBAs. First, we prove that it can effectively measure similarity in classic, and secondly, we implement it on quantum circuits to realize the measurement of BBA-QC.

\begin{definition}[FB inner product]\cite{zhou2021handling}\rm{
Given $2$ BBAs $m_1$ and $m_2$ under an $n$-element FoD $\Theta$, their fractal-based belief inner product is
\begin{equation}
    Inner_{\mathrm{FB}}(m_1,m_2)=\sum_{F\subseteq\Theta}\frac{m_\mathrm{F1}(F)m_\mathrm{F2}(F)}{||m_\mathrm{F1}||_2||m_\mathrm{F2}||_2},
\end{equation}
which also can be written as three steps:
\begin{itemize}
    \item Generate FBBA: $\boldsymbol{m_{\mathrm{F}1}'}=\boldsymbol{M}_\mathrm{F}\cdot \boldsymbol{m_1'}$ and $\boldsymbol{m_{\mathrm{F}2}'}=\boldsymbol{M}_\mathrm{F}\cdot \boldsymbol{m_2'}$. We only consider the normal BBA, so the $\boldsymbol{M}_\mathrm{F}=\begin{bmatrix} 1 &0\\0&M^*_\mathrm{F} \end{bmatrix}$, and $\boldsymbol{M^*}_\mathrm{F}(F,G)=\frac{1}{2^{G}-1}$ iff $F\subseteq G$ and $F$,$G$ are non-empty subsets of $\Theta$.
    \item Normalized FBBA: $\boldsymbol{m_{\mathrm{F}1\_\mathrm{N}}}=\frac{\boldsymbol{m_{\mathrm{F}1}}}{||\boldsymbol{m_{\mathrm{F}1}}||_2}$ and $\boldsymbol{m_{\mathrm{F}2\_\mathrm{N}}}=\frac{\boldsymbol{m_{\mathrm{F}2}}}{||\boldsymbol{m_{\mathrm{F}2}}||_2}$.
    \item Inner product: $Inner_{\mathrm{FB}}(m_1,m_2)=\boldsymbol{m_{\mathrm{F}1\_\mathrm{N}}}\cdot \boldsymbol{m_{\mathrm{F}2\_\mathrm{N}}'}$.
\end{itemize}}
\end{definition}

FB inner product has only weak structural properties \cite{jousselme2012distances} based on the above definition , i.e. the expression is related to cardinality $F$. However, $M^*_\mathrm{F}$ is an upper triangular matrix and $M^{*'}_\mathrm{F}M^*_\mathrm{F}=D$, where $D(F,G)\neq 0$ iff $D\cap G\neq\varnothing$ and its value is related to the structure of focal sets. Hence, the measures based on fractal-based transformation meet strong structural property and structural dissimilarity.

\begin{example}\label{efb}\rm{
For a $10$-element FoD, two BBAs are $m_1=\{m_1(A)=0.8,m_1(\theta_7)=0.05, m_1(\theta_2\theta_3\theta_4)=0.05, m(\Theta)=0.1\}$ and $m_2=\{m(\theta_1\theta_2\theta_3\theta_4\theta_5)=1\}$. When the $A$ from $\{\theta_1\}$ to $\{\Theta\}$, the change trends of evidence distance $d_\mathrm{BBA}$, FB inner product $Inner_{\mathrm{FB}}$, classical fidelity $Fid$, Euclidean distance $d$, and evidence inner product $Inner_\mathrm{BBA}$ \footnote{$Inner(m_1,m_2)=\boldsymbol{m_1}D_\mathrm{E}\boldsymbol{m_2'}$\cite{jousselme2012distances}} are shown in Figure \ref{toy_example}.}

\begin{figure}[htbp!]
\centering
\includegraphics[width=0.7\textwidth]{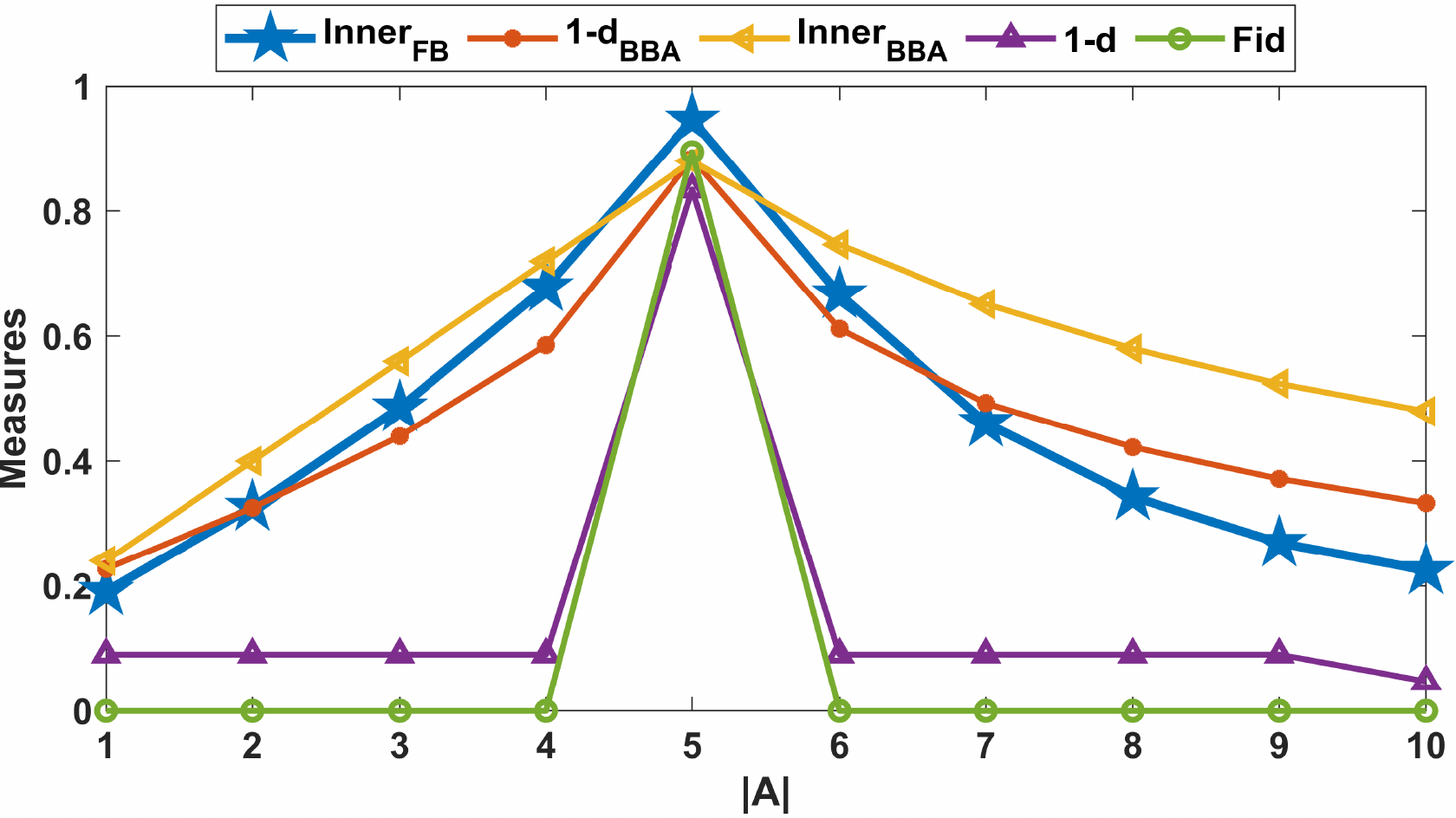}
\caption{The change trend of BBA measures. In order to show them more intuitive, the $d_\mathrm{BBA}$ and $d$ are $1-d_\mathrm{BBA}$ and $1-d$ respectively, which can be seen as a similarity measure.}
\label{toy_example}
\end{figure}

\end{example}

Figure \ref{toy_example} proves that FB inner product has a reasonable trend in measuring similarity when the focal set frame is changed.

\subsubsection{FB inner product on quantum circuits}
\begin{figure*}[htbp!]
\centering
\includegraphics[width=0.75\textwidth]{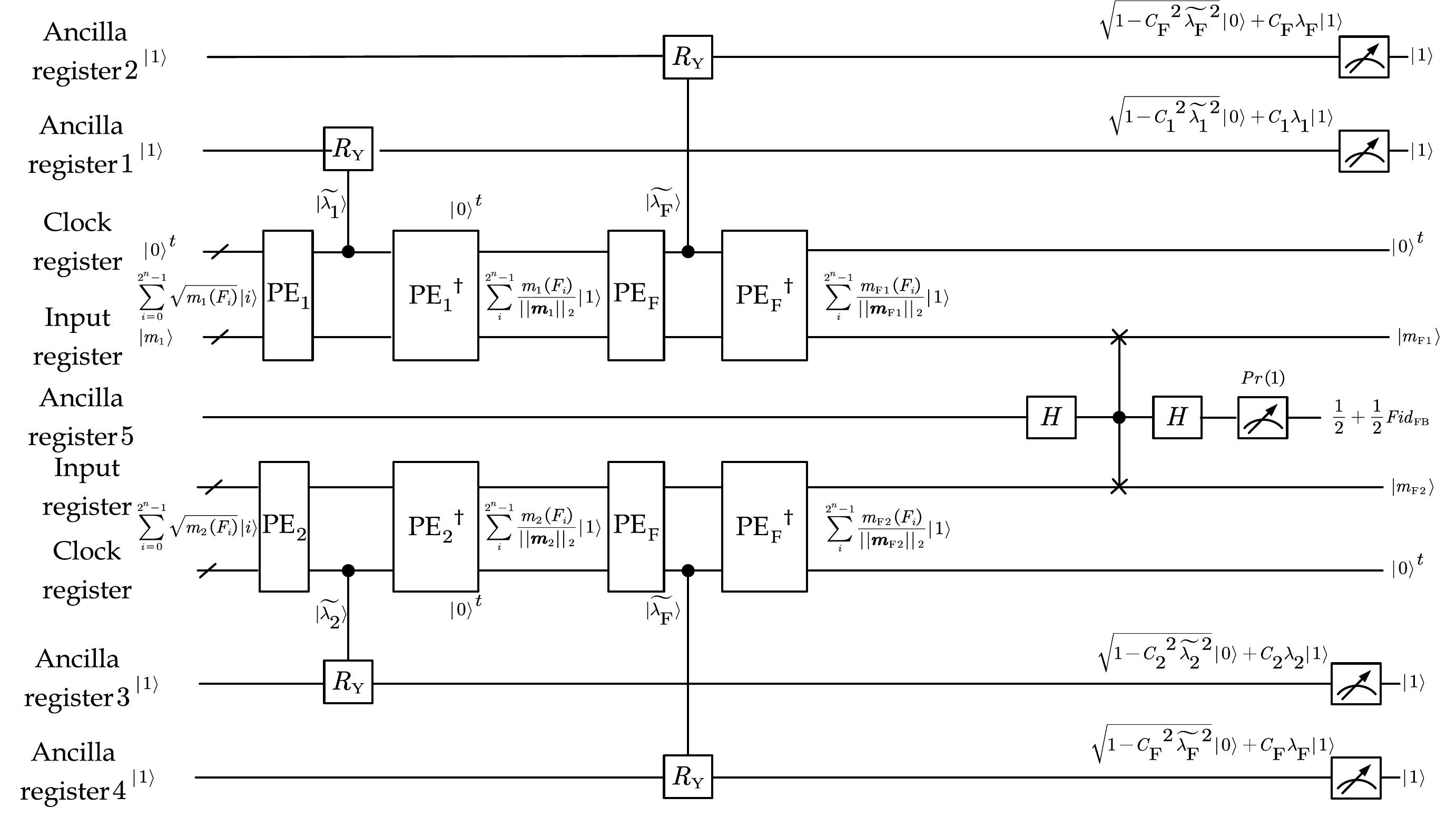}
\caption{Measure the similarity of BBA-QC by FB inner product.}
\label{FFB-QC}
\end{figure*}

Since the BBA-QC is $\ket{m}=\sum_{F_i\subseteq\Theta}\sqrt{m(F_i)}\ket{i}$, the $m_\mathrm{F}$ cannot be implemented by directly performing the $M_\mathrm{F}$ operation. Hence, we need to change $\ket{m}$ into $\sum_{F_i\subseteq\Theta}\frac{m(F_i)}{||\boldsymbol{m'}||_2}\ket{i}$ through the diagonal matrix $diag(\sqrt{\boldsymbol{m}})$. Due to the feature that the quantum state amplitude is normalized, after executing $\boldsymbol{M}_\mathrm{F}$, $\sum_{F_i\subseteq\Theta}\frac{m_F(F_i)}{||\boldsymbol{m_F'}||_2}\ket{i}$ can be obtained directly. Then the swap test is used to measure the similarity and out put the result. The specific quantum circuit is shown in Figure \ref{FFB-QC}. $9$ registers are needed to complete the similarity measure. When the measurements of ancilla registers $1-4$ are $\ket{1}$, measuring ancilla register $5$ constant times can extract the similarity.

\subsection{MEoB-based implement belief functions}

In Section \ref{QBF}, we utilize C-NOT gates to extract belief functions from BBA-QC. According to Equation \ref{belief_m} and MEoB, we can implement a quantum state including partial information of belief functions. The circuit is similar with fractal-based transformation

\begin{equation}
\begin{aligned}
    \ket{m}=\sum_{i=0}^{2^n-1}\sqrt{m(F_i)}\ket{i}\stackrel{diag(\sqrt{m})}{\longrightarrow} \ket{\widehat{m}}=\sum_{i=0}^{2^n-1}\frac{m(F_i)}{||\boldsymbol{m}||_2}\stackrel{\boldsymbol{M}_\mathrm{Bel};\boldsymbol{M}_\mathrm{Pl}\boldsymbol{M}_\mathrm{q}}{\longrightarrow}\text{output state},
\end{aligned}
\end{equation}
and the output state are
\begin{equation}\label{bc-qc2}
\begin{aligned}
        &\ket{\widehat{Bel}}=\sum_{i=1}^{2^n-1}\frac{Bel(F_i)}{||\boldsymbol{Bel}||_2}\ket{i}; \ket{\widehat{Pl}}=\sum_{i=0}^{2^n-1}\frac{Pl(F_i)}{||\boldsymbol{Pl}||_2}\ket{i};\ket{\widehat{q}}=\sum_{i=0}^{2^n-1}\frac{q(F_i)}{||\boldsymbol{q}||_2}\ket{i};
\end{aligned}
\end{equation}
where $\ket{\widehat{\cdot}}$ means that not all information can be extracted by measurements because of normalization. The sum of belief functions is not certain, so they can not be recovered in classical. Therefore, this method can only be applied to continue operating belief functions on quantum circuits, rather than extracting it precisely.

\subsection{Evidence Combination Rules of BBA-QC}

\begin{figure*}[htbp!]
\centering
\includegraphics[width=0.95\textwidth]{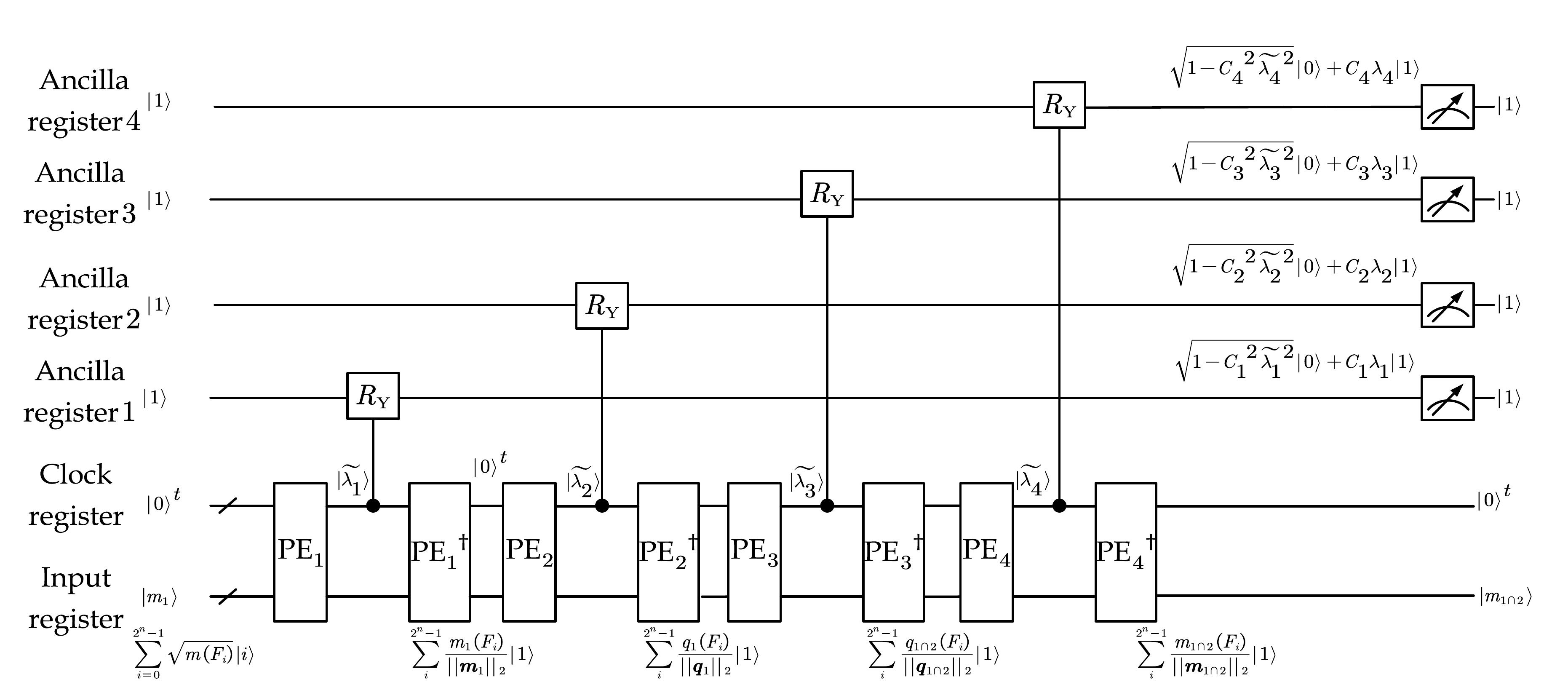}
\caption{CCR on quantum circuits. When the measurements of ancilla registers $1-4$ are $\ket{1}$, the input register stores the goal state.}
\label{QCCR}
\end{figure*}

Given to BBAs $m_1$ and $m_2$, the CCR and DCR can be realized by matrix operations \cite{smets2002application} $\boldsymbol{m_{1\circledtiny{$\cap$}2}'}=\boldsymbol{S_{m_1}}\cdot\boldsymbol{m_2'}$ and $\boldsymbol{m_{1\circledtiny{$\cup$}2}'}=\boldsymbol{G_{m_1}}\cdot\boldsymbol{m_2'}$, where $\boldsymbol{S_{m}}$ and $\boldsymbol{G_{m}}$ are upper triangular matrix and lower triangular matrix respectively. Though MEoB can accelerate the above operations on quantum circuits, $\boldsymbol{S_{m}}$ and $\boldsymbol{G_{m}}$ are depended on  specific pre-fused BBAs. Hence, the process of transformation BBA into a matrix also needs to be considered in the entire algorithm. According to Equation \ref{ccr} and \ref{dcr},they can be implemented by
\begin{equation}
\boldsymbol{S_{m}}=\boldsymbol{M}_\mathrm{q}^{-1}diag(q)\boldsymbol{M}_\mathrm{q};~~\boldsymbol{G_m}=\boldsymbol{M}_\mathrm{b}^{-1}diag(b)\boldsymbol{M}_\mathrm{b},
\end{equation}
where $\boldsymbol{M_b}=\begin{bmatrix}1&\boldsymbol{M}_\mathrm{Bel}\\1&1\end{bmatrix}$. For different BBAs, only the diagonal matrix is different, which can be implemented using the circuit shown in Figure \ref{BF-QC}. Therefore, utilizing $3$ metrics can realize the evidence CRs on the quantum circuit. Figure \ref{QCCR} shows the quantum circuits of CCR, and the implementation process is
\begin{equation}
    \ket{m_1}\stackrel{diag(\sqrt{m_1})}{\longrightarrow}\ket{\widehat{m_1}}\stackrel{M_\mathrm{q}}{\longrightarrow}\ket{\widehat{q_1}}\stackrel{diag(q_2)}{\longrightarrow}\ket{\widehat{q_{1\circledtiny{$\cap$}2}}}\stackrel{M_\mathrm{q}^{-1}}{\longrightarrow}\ket{\widehat{m_{1\circledtiny{$\cap$}2}}}.
\end{equation}
The output state $\ket{\widehat{m_{1\circledtiny{$\cap$}2}}}=\frac{m_{1\circledtiny{$\cap$}2}(F_i)}{||\boldsymbol{m}_{1\circledtiny{$\cap$}2}||_2}\ket{i}$ can be extracted by measurements. Though it is the normalized result, we can recover it based on $m_{F\subseteq\Theta}(F)=1$. In addition, due to the normalization step, this method cannot be directly applied to DRC, and only can calculate according to Equation \ref{DRC} after executing CCR.

\subsection{Probability transformation of BBA-QC}
\subsubsection{Pignistic probability transformation}
Similar with the evidence CRs, PPT also can be realized by matrix operation \cite{smets2002application}. We extent the $BetP$ to the whole power set $BetM(F)=\sum_{G\subseteq\Theta}|F\cap G|m(G)/|G|(1-m(\varnothing))$, where $BetP=\{BetM(\theta_1),\cdots, BetM(\theta_n)\}$. According to the matrix operations, $\boldsymbol{BetM'}=\boldsymbol{M}_\mathrm{Bet}\cdot \boldsymbol{m'}=\boldsymbol{Cred}\cdot \boldsymbol{D} \cdot \boldsymbol{m'}$, where $\boldsymbol{Cred}(F,G)=|F\cap G|$, $D(F,G)=\begin{cases}
\frac{1}{|F|} & \textit{if $F=G$} \\
0 & \text{Others}
\end{cases}$, and $\boldsymbol{M}_\mathrm{Bet}=\boldsymbol{Cred}\cdot \boldsymbol{D}$ is an Hermitian matrix. Based on the above, we can implement $BetM$ on quantum circuits
\begin{equation}
    \ket{m}\stackrel{diag(\sqrt{m})}{\longrightarrow}\ket{\widehat{m}}\stackrel{M_\mathrm{Bet}}{\longrightarrow}\ket{\widehat{BetM}}.
\end{equation}
Similar with the evidence CRs on quantum circuits, though the output state is a normalized result, $BetM$ also can be recovered based on $\sum_{\theta_i\in \Theta}BetP(\theta_i)=1$.
\subsubsection{Plausibility probability transformation}
The key of PMT is to implement the $Pl(\theta)$ from $m$. According to Equation \ref{qbelfe1} and \ref{bc-qc2}, there are $2$ methods can extract information $Pl(\theta)$ from $\ket{m}$. Obviously, the methods of Equation \ref{qbelfe1} is more convenient. $\mathcal{O}(n)$ complexity is needed to extract complete $\{Pl(\theta_1)\cdots Pl(\theta_n)\}$ on quantum circuits, and $\mathcal{O}(2n)$ complexity to normalize them in classical.
\subsection{Discussion}

In this section, we propose methods to implement common operations of DST on quantum circuits. Table \ref{discussion} shows the difference between them in quantum and classical operations. For a fair comparison, the algorithms in classical and quantum are not one-to-one, but chooses the lowest one with the same effect. For example, in similarity measurements, the classical algorithm is the evidence distance, and the quantum is the FB inner product. In the evidence CRs, the classical are Equation \ref{ccr1}, and the quantum is their matrix operations. Information loss refers to whether the results extracted after measurements are the goal information, and recovery refers to whether the lost information can be recovered in classical operations. After comparison, it can be found that methods other than PPT can be accelerated on quantum circuits, which illustrates the feasibility of reducing the computational complexity of the DST operations on quantum computers.

\begin{table*}[htbp!]
    \centering
    \small
     \caption{The comparison of classical and quantum operations in DST.}
    \label{discussion}
    \begin{tabular}{c|c|c|c|c}
      Methods   & Complexity (classical) & Complexity (QC) & \tabincell{c}{Information\\ Loss} & Recover \\
      \hline
      Similarity &$\mathcal{O}(4^n+2^n)$&$\mathcal{O}(2n(2^n+\frac{\log\epsilon}{\log\log\epsilon})+poly(2^n))$&\XSolid&-\\
      \tabincell{c}{Belief Functions \\(total)}&$\mathcal{O}(4^n)$&$\mathcal{O}((2^n))$&\XSolid&-\\
      \tabincell{c}{Belief Functions \\(partial)}&$\mathcal{O}(4^n)$&$\mathcal{O}(n(2^n+\frac{\log\epsilon}{\log\log\epsilon}))$&\Checkmark&\XSolid\\
      Evidence CRs &$\mathcal{O}(4^n)$&$\mathcal{O}(4n(2^n+\frac{\log\epsilon}{\log\log\epsilon}))$&\Checkmark&\Checkmark\\
      PPT &$\mathcal{O}(2^n)$&$\mathcal{O}(n(2^n+\frac{\log\epsilon}{\log\log\epsilon}))$&\Checkmark&\Checkmark\\
      PTM &$\mathcal{O}(n2^n+n)$&$\mathcal{O}(3n)$&\XSolid&-\\
    \end{tabular}
\end{table*}

\section{Conclusion}
\label{con}
In order to solve the problem of the power exponential explosion of DST algorithms on classical computers, this paper encodes the BBA into quantum states and accelerates the common operations of BBA on quantum computers. Even if the quantum states are orthogonal, this cannot correspond to the incompletely mutually exclusive focal sets in a physical sense, we still can use the characteristics of quantum computing to speed up the operations. The characteristic of quantum computation is $n$ qubits can represent $2^n$ data at the same time, which is consistent with utilizing $2^n$ mass functions to describe $n$ elements in DST. We encode the BBA into the quantum state, which allows qubits to directly control the corresponding elements' mass functions. In classical, the order of mass functions is often considered when dealing with BBA. However, in quantum computation, the qubits operation is equivalent to directly changing the belief functions, which makes the DST has great application prospects in quantum computation.

Regarding BBA-QC, the specific operations of the paper are as follows: (1) We propose the specific algorithm for preparing BBA-QC, and prove that the extraction of BBA-QC can be exponentially accelerated compared to the classical. (2) We propose the implementation of accurate BF-QC through the C-NOT gates, which also achieves an exponential speedup compared to classical calculations. (3) Based on the HHL algorithm, we propose MEoB algorithm that operates BBA with a matrix on quantum circuits. (4) For common operations in DST, we give the corresponding implementation on quantum circuits based on MEoB. While achieving the same effect, we find that similarity measure, evidence CRs, and PMT can achieve satisfactory speed-ups on quantum circuits.

In the following research, we will mainly focus on the following $3$ aspects (1) Although some operations of DST have been accelerated on quantum circuits in this paper, they are all based on MEoB algorithm, when it is difficult to transform them into matrix operations in classical, they can not be implemented on quantum circuits. Therefore, in future research, we plan to utilize variation quantum algorithms to process BBA-QC, so that more operations can be implemented on quantum circuits. (2) When using quantum computation to deal with BBA, we find that BBA-QC can realize that one qubit controls one element, i.e., some qubits are operated, the corresponding focal sets containing these elements can be operated simultaneously. Hence, we plan to explore more quantum advantages besides complexity. (3) For other uncertainty theories, such as Random Permutation Set \cite{deng2022RPS} and General Credal Sets , how to operate them on quantum circuits is also an open issue worthy of discussion.


\section*{Acknowledgment}

The work is partially supported by National Natural Science Foundation of China (Grant No. 61973332 and No. 61801459).

 \bibliographystyle{elsarticle-num}
 \bibliography{cas-refs}





\end{document}